\documentclass[showpacs,showkeys,preprintnumbers,eqsecnum,prd,twocolumn]{revtex4}  

\usepackage{graphicx}

\def\beq{\begin{equation}}
\def\eeq{\end{equation}}
\def\rmd{{\rm d}}

\begin{document}

\title{Dynamics of quadrupolar bodies in a Schwarzschild spacetime}

\author{Donato Bini}
  \affiliation{
Istituto per le Applicazioni del Calcolo 
``M. Picone,'' CNR, I-00185 Rome, Italy\\
ICRA, University of Rome ``La Sapienza,'' 
00185 Rome, Italy\\
INFN, Sezione di Firenze, Polo Scientifico, Via Sansone 1, 
50019 Sesto Fiorentino, Florence, Italy
}

\author{Andrea Geralico}
  \affiliation{Physics Department and ICRA, ``Sapienza" University of Rome, I-00185 Rome, Italy}

\begin{abstract}
The dynamics of extended bodies endowed with multipolar structure up to the mass quadrupole moment is investigated in the Schwarzschild background according to the Dixon's model, extending previous works.
The whole set of evolution equations is numerically integrated under the simplifying assumptions of constant frame components of the quadrupole tensor and that the motion of the center of mass be confined on the equatorial plane, the spin vector being orthogonal to it.
The equations of motion are also solved analytically in the limit of small values of the characteristic length scales associated with the spin and quadrupole with respect to the background curvature characteristic length.
The results are qualitatively and quantitatively different from previous analyses involving only spin structures.
In particular, the presence of the quadrupole turns out to be responsible for the onset of a non-zero spin angular momentum, even if initially absent.
\end{abstract}


\maketitle

\section{Introduction}

Deviations from geodesic motion of extended bodies in a given background spacetime due to their multipolar structure were first studied by Mathisson and Papapetrou \cite{math37,papa51}, who obtained a set of evolution equations for both linear and angular momentum of the body.
This model was later improved by several authors, including Tulczyjew \cite{tulc59}, Dixon \cite{dixon64,dixon69,dixon70,dixon73,dixon74,Dixon2} and Ehlers and Rudolph \cite{ehlers77}, whose work lead to a deeper understanding of the role of the multipolar  (quadrupolar, in detail)  structure of the body.
Since then the model equations have been treated by different approximation schemes both analytically and numerically in astrophysically relevant background spacetimes. 
Examples of purely numerical studies of the full nonlinear equations can be found in Refs. \cite{maeda,ver,hartl1,hartl2,sem99,singh}.
The existence of analytic solutions is only allowed in special situations  which are in general too restrictive to yield a complete description of the nongeodesic motion induced by the structure of the body; for instance, by constraining the path along Killing trajectories in highly symmetric spacetimes, e.g., circular orbits \cite{bdfg2004a,bdfg2004b,bdfgj2005a,bdfgj2005b,bdfgj2005c,bdfgj2005d,bgj2006}, also in the ultrarelativistic regime \cite{ply}.  Finally, for bodies with structure up to the octupole momentum an action principle formulation of the dynamics has been studied, with applications limited to gravitational phase shift \cite{anandan}.

In the present paper we study the dynamics of extended bodies endowed with both dipolar and quadrupolar structure according to the so called Mathisson-Papapetrou-Dixon (MPD) model exploring the fully relativistic content of the model itself.
The background is chosen to be the Schwarzschild spacetime and the motion is assumed to be confined on the equatorial plane, the spin vector being orthogonal to it.
The dynamics is well specified in the case of a purely spinning object only, because the MPD equations directly determine the motion as well as the spin evolution itself.
On the contrary, in the MPD model there are no evolution equations for the quadrupole as well as higher multipoles, so their evolution is completely free, depending only on the considered body.
The model simply implies the addition to the equations of quadrupolar force and torque as source terms which modify the evolution of both the linear and angular momentum of the body.
Therefore, one has to supply the structure of the body as an external information, like the constitutive equations in classical continuum dynamics.
This fact represents a limitation to the model itself which allows for many different approaches. 
In our analysis we assume that all unspecified quantities describing the shape of the body are constant in the frame associated with the 4-momentum of the body itself and that are known as intrinsic properties of the matter under consideration.  
It would be of great interest to extend this analysis to systems with varying quadrupolar structure and emitting gravitational waves without perturbing significantly the background spacetime. 
Usually variable quadrupole moment is generated in a test astronomical body of mass $m$ because of the tides produced by the central source of mass $M\gg m$. Here the net gravitational radiation associated with the motion of $m$ is due to its orbit around $M$, the time varying tides and the interference between these two \cite{mash1,mash2}. 
We deserve such an investigation to future works.

Quadrupolar effects on the motion of extended bodies could be very important in many realistic astrophysical situations.
Consider, for instance, an extended body moving around a central source, like binary pulsar systems orbiting the Galactic Center supermassive black hole (Sgr A$^*$) \cite{falcke,lyne}.
Measuring the period of revolution (known from observations) will provide an estimate of the quantities determining the quadrupolar structure of the body, if its spin is known \cite{quad_schw,quad_kerr}. On the other hand, the complete knowledge of the internal structure of the body will allow to estimate the period of revolution. 
Such analysis can also be extended to other objects of astrophysical interest, e.g., ordinary or neutron stars, around Sgr A$^*$. The interest to study orbits close to Sgr A$^*$ relies on the increasing accuracy in sub-milli-arcsecond astrometry by the near-infrared detectors \cite{eise} and on the potentiality of the next-generation radiotelescopes, e.g., the Square Kilometer Array (SKA) \cite{ska}, to identify some of the $10^4$ compact objects orbiting within 1 pc around Sgr A$^*$ \cite{muno}.
Applications to cosmological spacetimes have also been considered (see, e.g., Ref. \cite{harte}).

\section{Dixon's model and basic equations}

Consider an extended body endowed with structure up to the quadrupole, following the description due to Dixon \cite{dixon64,dixon69,dixon70,dixon73,dixon74}.
In the quadrupole approximation Dixon's equations are
\begin{eqnarray}
\label{papcoreqs1}
\frac{{\rm D}P^{\mu}}{\rmd \tau} & = &
- \frac12 \, R^\mu{}_{\nu \alpha \beta} \, U^\nu \, S^{\alpha \beta}
-\frac16 \, \, J^{\alpha \beta \gamma \delta} \, \nabla^\mu R_{\alpha \beta \gamma \delta}
\nonumber\\
& \equiv & F^\mu_{\rm (spin)} + F^\mu_{\rm (quad)} \,,
\\
\label{papcoreqs2}
\frac{{\rm D}S^{\mu\nu}}{\rmd \tau} & = & 
2 \, P^{[\mu}U^{\nu]}+
\frac43 \, J^{\alpha \beta \gamma [\mu}R^{\nu]}{}_{\gamma \alpha \beta}
\nonumber\\
&\equiv & D^{\mu \nu}_{\rm (spin)} + D^{\mu \nu}_{\rm (quad)} \,,
\end{eqnarray}
where $P^{\mu}=m u^\mu$ (with $u \cdot u = -1$) is the total 4-momentum of the particle, $S^{\mu \nu}$ is a (antisymmetric) spin tensor, $J^{\alpha\beta\gamma\delta}$ is the quadrupole moment of the stress-energy tensor of the body, and $U$ is the timelike unit tangent vector of the \lq\lq centre of mass line'' used to make the multipole reduction, parametrized by the proper time $\tau$.
In order the model to be mathematically correct certain additional conditions should be imposed \cite{tulc59,dixon64}
\beq
\label{tulczconds}
S^{\mu\nu}u{}_\nu=0\,.
\eeq
Consequently, the spin tensor can be fully represented by a spatial vector (with respect to $u$),
\beq
S(u)^\alpha=\frac12 \eta(u)^\alpha{}_{\beta\gamma}S^{\beta\gamma}=[{}^{*_{(u)}}S]^\alpha\,,
\eeq
where 
\beq
\eta(u)_{\alpha\beta\gamma}=\eta_{\mu\alpha\beta\gamma}u^\mu
\eeq
is the spatial (with respect to $u$) unit volume 3-form with $\eta_{\alpha\beta\gamma\delta}=\sqrt{-g} \epsilon_{\alpha\beta\gamma\delta}$ the unit volume 4-form and $\epsilon_{\alpha\beta\gamma\delta}$ ($\epsilon_{0123}=1$) is the Levi-Civita alternating symbol. As standard, hereafter we denote the spacetime dual of a tensor (built up with $\eta_{\alpha\beta\gamma\delta}$) by a $^*$, whereas the 
spatial dual of a spatial tensor with respect to $u$ (built up with $\eta(u)_{\alpha\beta\gamma}$) by $^{*_{(u)}}$.
It is also useful to introduce the magnitude $s\ge0$ of the spin vector
\beq
\label{sinv}
s^2=S(u)^\beta S(u)_\beta = \frac12 S_{\mu\nu}S^{\mu\nu}\,, 
\eeq
which is in general not constant along the trajectory of the extended body. 

Note that, in Eqs. (\ref{papcoreqs1}) and (\ref{papcoreqs2}), the spin force and torque depend on both the world line of multipole reduction $U$ and the generalized momentum unit vector $u$, but the same is not true for the quadrupolar force and torque, which only depend on the spacetime quantity $J^{\alpha\beta\gamma\delta}$ and the background geometry.

The tensor $J^{\alpha\beta\gamma\delta}$ has the same algebraic symmetries as the Riemann tensor, i.e.,
\beq
J^{\alpha\beta\gamma\delta}=J^{[\alpha\beta][\gamma\delta]}=J^{\gamma\delta\alpha\beta}\,,\qquad
J^{[\alpha\beta\gamma]\delta}=0\,,
\eeq
leading to 20 independent components.
Using standard spacetime splitting techniques it can be reduced to the following form
\begin{eqnarray}
\label{deco_bar_u}
J^{\alpha\beta\gamma\delta}&=&\Pi(\bar u)^{\alpha\beta\gamma\delta}+2\bar u^{[\alpha}\pi(\bar u)^{\beta]\gamma\delta}
+2\bar u^{[\gamma}\pi(\bar u)^{\delta]\alpha\beta}\nonumber\\
&&-4\bar u^{[\alpha}Q(\bar u)^{\beta][\gamma}\bar u^{\delta]}\,,
\end{eqnarray}
where $Q(\bar u)^{\alpha\beta}=Q(\bar u)^{(\alpha\beta)}$ represents the quadrupole moment of the mass distribution as measured by an observer with $4$-velocity $\bar u$.
[Note that our mass quadrupole moments $Q(\bar u)$ and $\pi(\bar u)$ differ for the analogous quantities introduced by Ehlers and Rudolph \cite{ehlers77} by numerical factors: $Q(\bar u)=(3/4)Q_{\rm ER}(\bar u)$ and $\pi(\bar u)=-(1/2)\pi_{\rm ER}(\bar u)$.
Our choice is motivated by the use of a representation of $J$ formally analogue to the Riemann tensor, see below.]
Similarly $\pi(\bar u)^{\alpha\beta\gamma}=\pi(\bar u)^{\alpha [\beta\gamma]}$ (with the additional property $\pi(\bar u)^{[\alpha\beta\gamma]}=0$) and 
$\Pi(\bar u)^{\alpha\beta\gamma\delta}=\Pi(\bar u)^{[\alpha\beta][\gamma\delta]}$ are essentially the body's momentum and stress quadrupole.
Moreover the various fields $Q(\bar u)^{\alpha\beta}$, $\pi(\bar u)^{\alpha\beta\gamma}$ and $\Pi(\bar u)^{\alpha\beta\gamma\delta}$ are all spatial with respect to $\bar u$, i.e., give zero after any contraction by $\bar u$ 
\beq
{\bar u}{}_\alpha\Pi({\bar u})^{\alpha \beta\gamma\delta}={\bar u}{}_\alpha\pi({\bar u})^{\alpha \beta\gamma}={\bar u}{}_\alpha Q({\bar u})^{\alpha \beta}=0\,.
\eeq
As stated above, the number of independent components of $J^{\alpha\beta\gamma\delta}$ is 20: 6 in $Q(\bar u)^{\alpha\beta}$, 6 in $\Pi(\bar u)^{\alpha\beta\gamma\delta}$ and 8 in $\pi(\bar u)^{\alpha\beta\gamma}$. 
The representation (\ref{deco_bar_u}) of $J$ is analogous to the standard $1+3$ representation of the Riemann tensor in terms of its electric, magnetic and mixed parts defined as
\begin{eqnarray}
E(\bar u)_{\alpha\beta}&=&R_{\alpha\mu\beta\nu}\bar u^\mu \bar u^\nu\,,\nonumber\\
H(\bar u)_{\alpha\beta}&=&-[R^*]_{\alpha\mu\beta\nu}\bar u^\mu \bar u^\nu\,,\nonumber\\
F(\bar u)_{\alpha\beta}&=&[{}^*R^*]_{\alpha\mu\beta\nu}\bar u^\mu \bar u^\nu\,,
\end{eqnarray}
so that in vacuum (where the mixed part $F(\bar u)_{\alpha\beta}=-E(\bar u)_{\alpha\beta}$) one has
\begin{eqnarray}
\label{rieman_deco}
R^{\alpha\beta\gamma\delta}\!\!&=&\!\!-\eta(\bar u)^{\alpha \beta \mu}\eta(\bar u)^{\gamma\delta\nu}E(\bar u)_{\mu\nu}
+2\bar u^{[\alpha}H(\bar u)^{\beta]}{}_\sigma \eta(\bar u)^{\sigma \gamma\delta}\nonumber\\
&&+2\bar u^{[\gamma}H(\bar u)^{\delta]}{}_\sigma \eta(\bar u)^{\sigma \alpha\beta}
-4\bar u^{[\alpha}E(\bar u)^{\beta][\gamma}\bar u^{\delta]}\,.\nonumber\\
\end{eqnarray}

Taking the right dual of (\ref{deco_bar_u}) gives
\begin{eqnarray}
\label{deco_bar_udual}
[J^*]^{\alpha\beta\mu\nu}&=&\frac12J^{\alpha\beta\gamma\delta}\eta_{\gamma\delta}{}^{\mu\nu}\nonumber\\
&=&[\Pi(\bar u)^*]^{\alpha\beta\mu\nu}+2\bar u^{[\alpha}[\pi(\bar u)^*]^{\beta]\mu\nu}\nonumber\\
&&+\eta(\bar u)_\delta{}^{\mu\nu}\pi(\bar u)^{\delta\alpha\beta}
+2\eta(\bar u)_{\gamma}{}^{\mu\nu}\bar u^{[\alpha}Q(\bar u)^{\beta]\gamma}\,,\nonumber\\
\end{eqnarray}
while taking both the left and right dual leads to
\begin{eqnarray}
\label{deco_bar_udualdual}
[{}^*J^*]^{\rho\sigma\mu\nu}&=&\frac12\eta_{\alpha\beta}{}^{\rho\sigma}[J^*]^{\alpha\beta\mu\nu}
=\frac14 \eta_{\alpha\beta}{}^{\rho\sigma}J^{\alpha\beta\lambda\delta} \eta_{\lambda\delta}{}^{\mu\nu} \nonumber\\
&=&[{}^*\Pi(\bar u)^*]^{\rho\sigma\mu\nu}
+2\eta(\bar u)_\beta{}^{\rho\sigma}[\pi(\bar u)^*]^{\beta\mu\nu}\nonumber\\
&&+\eta(\bar u)_\delta{}^{\mu\nu}[\pi(\bar u)^*]^{\delta\rho\sigma}\nonumber\\
&&+2\eta(\bar u)_{\gamma\delta}{}^{\mu}\eta(\bar u)_{\beta}{}^{\rho\sigma}Q(\bar u)^{\beta\gamma}\,,
\end{eqnarray}
where $\eta(\bar u)^{\alpha\beta\gamma}=\bar u_\sigma \eta^{\sigma \alpha\beta\gamma}$.
The above decomposition (\ref{deco_bar_u}) thus becomes
\begin{eqnarray}
\label{deco_bar_u2}
J^{\alpha\beta\gamma\delta}&=&\eta(\bar u)^{\alpha \beta \mu}\eta(\bar u)^{\gamma\delta\nu}M(\bar u)_{\mu\nu}
+2\bar u^{[\alpha}W(\bar u)^{\beta]}{}_\sigma \eta(\bar u)^{\sigma \gamma\delta}\nonumber\\
&&+2\bar u^{[\gamma}W(\bar u)^{\delta]}{}_\sigma \eta(\bar u)^{\sigma \alpha\beta}
-4\bar u^{[\alpha}Q(\bar u)^{\beta][\gamma}\bar u^{\delta]}\,,\nonumber\\
\end{eqnarray}
so that
\begin{eqnarray}
\label{Jsplit_baru}
Q(\bar u)_{\alpha\beta}&=&J_{\alpha\mu\beta\nu}\bar u^\mu \bar u^\nu\,,\nonumber\\ 
W(\bar u)_{\alpha\beta}&=&-[J^*]_{\alpha\mu\beta\nu}\bar u^\mu \bar u^\nu\,,\nonumber\\
M(\bar u)_{\alpha\beta}&=&[{}^*J^*]_{\alpha\mu\beta\nu}\bar u^\mu \bar u^\nu\,,
\end{eqnarray}
and
\begin{eqnarray}
\Pi(\bar u)^{\alpha\beta\gamma\delta}&=&\eta(\bar u)^{\alpha \beta \mu}\eta(\bar u)^{\gamma\delta\nu}M(\bar u)_{\mu\nu}\,,\nonumber\\
\pi(\bar u)^{\beta\gamma\delta}&=&W(\bar u)_{\beta\lambda}\eta(\bar u)_\lambda{}^{\gamma\delta}\,.
\end{eqnarray}
When the observer $\bar u=u$, i.e., when the observer is at rest with respect to the body, the spatial tensors $Q(u)^{\alpha\beta}$, $\pi(u)^{\alpha\beta\gamma}$ and $\Pi(u)^{\alpha\beta\gamma\delta}$, or equivalently the spatial tensors $Q(u)_{\alpha\beta}$, $W(u)_{\alpha\beta}$ and $M(u)_{\alpha\beta}$, have an intrinsic meaning.

The system of equations (\ref{papcoreqs1})--(\ref{papcoreqs2}) has been solved analytically in Refs. \cite{quad_schw,quad_kerr} in special situations, i.e., under the simplifying assumption of constant frame components (with respect to a natural orthonormal frame) of both the spin and the quadrupole tensor, obtaining the kinematical conditions to be imposed to the particle's structure in order the orbit of the particle itself be circular and confined on the equatorial plane of a Schwarzschild and Kerr black holes. 
Furthermore, the total 4-momentum of the body was taken aligned with $U$ and the only contribution to the complete quadrupole moment $J^{\alpha\beta\gamma\delta}$ was assumed to stem from the mass quadrupole moment $Q(U)^{\alpha\beta}$, so that $\pi(U)^{\alpha\beta\gamma}=0=\Pi(U)^{\alpha\beta\gamma\delta}$.
The latter condition was also used in Refs. \cite{quad_gw_weak,quad_gw_ex}, where the \lq\lq reaction'' of an extended body to the passage of both a weak and an exact plane gravitational wave is discussed.
In the present paper we relax both these assumptions, by taking the quadrupole tensor in its completely general form and with $P$ having no relation a priori with $U$.
Nevertheless, we specialize our analysis to the case in which the quadrupole tensor has constant frame components with respect to the frame adapted to $P$ as the most natural and simplifying choice. 
According to the terminology introduced in Ref. \cite{ehlers77}, the extended body should be termed in this case as \lq\lq quasi-rigid.'' 
Other approaches (equally valid in the framework of the MPD model equations) assume the quadrupole tensor be directly related to the Riemann tensor, having the same symmetry properties.
For instance, a minimal choice consists in taking 
\beq
J^{\alpha\beta\gamma\delta}=kR^{\alpha\beta\gamma\delta}\,,
\eeq
with $k$ constant.
A more refined choice would imply instead the electric and magnetic parts of the quadrupole tensor proportional to the electric and magnetic parts respectively of the Riemann tensor (see Eq. (\ref{rieman_deco})), namely
one can take $Q(\bar u)=-M(\bar u)=c_1E(\bar u)$ and $W(\bar u)=c_2H(\bar u)$, so that the final decomposition of $J$ is 
\begin{eqnarray}
\label{JproptoR}
J^{\alpha\beta\gamma\delta}&=&-c_1\eta(\bar u)^{\alpha \beta \mu}\eta(\bar u)^{\gamma\delta\nu}E(\bar u)_{\mu\nu}\nonumber\\
&&+2c_2\bar u^{[\alpha}H(\bar u)^{\beta]}{}_\sigma \eta(\bar u)^{\sigma \gamma\delta}\nonumber\\
&&+2c_2\bar u^{[\gamma}H(\bar u)^{\delta]}{}_\sigma \eta(\bar u)^{\sigma \alpha\beta}\nonumber\\
&&-4c_1\bar u^{[\alpha}E(\bar u)^{\beta][\gamma}\bar u^{\delta]}\,,
\end{eqnarray}
with $c_1$ and $c_2$ constant.
A similar definition of the quadrupole tensor has been adopted, e.g., in Ref. \cite{steinhoff} (see also references therein) to study quadrupole deformation effects induced by the tidal field of a black hole on the motion of a spinning body. 
According to such an approach, the constants $c_1$ and $c_2$ are identified with the gravitoelectric-type and gravitomagnetic-type quadrupole tidal coefficients $\mu_2$ and $\sigma_2$, respectively, of a self-gravitating body, introduced in Ref. \cite{damnag} to study the response of neutron stars to external relativistic tidal fields.
One could also include in the definition (\ref{JproptoR}) of the quadrupole tensor also terms which are quadratic in spin, as discussed in Ref. \cite{steinhoff} (see also Ref. \cite{bailey} for an action principle approach to the dynamics of extended bodies and Ref. \cite{steinhoff2} for a review on the use of an extention of the Arnowitt-Deser-Misner canonical formalism to this context). 
For instance, one can consider the choice \cite{steinhoff2}
\beq
\label{Jspininduced}
J^{\alpha\beta\gamma\delta}=-4u^{[\alpha}Q(u)^{\beta][\gamma}u^{\delta]}\,,
\eeq
with
\beq
Q(u)=-\frac{C_Q}{m}[S^2]^{\rm (TF)}\,,
\eeq
where $C_Q$ is a constant and $[S^2]^{\rm (TF)}$ denotes the trace-free part of the square of the spin tensor, i.e., 
\begin{eqnarray}
[S^2]^{\rm (TF)}{}^{\alpha\beta}&=&S^{\alpha\mu}S_{\mu}{}^{\beta}-\frac13P(u)^{\alpha\beta}S_{\rho\sigma}S^{\sigma\rho}\nonumber\\
&=&S(u)^{\alpha}S(u)^{\beta}-\frac13s^2P(u)^{\alpha\beta}\nonumber\\
&=&[S(u)\otimes S(u)]^{\rm (TF)}{}^{\alpha\beta}\,,
\end{eqnarray}
where both the spin vector and the associated spin invariant have been used.
Special values of $C_Q$ have been given in Ref. \cite{steinhoff3}: for instance, in the case of a black hole one has $C_Q=1$ \cite{thorne}, whereas for neutron stars it depends on the equation of state and varies between 4.3 and 7.4 \cite{poisson}.  

However, none of these choices, even if very convenient from a computational point of view, has a transparent physical meaning ---at least a priori--- in the context of the MPD model, because the quadrupole tensor represents the matter only, and cannot be specified at all by the background in which the body moves.

We close this section with a few general comments.
A first remark concerns the fact that the quadrupole tensor components are not subjected to any algebraic constraint, differently from what happens in the case of the spin tensor, whose electric components with respect to the proper frame of the body are killed by the Dixon-Tulczyjew conditions (\ref{tulczconds}), i.e., by some additional conditions ensuring the model be mathematically well posed. This discrepancy can be actually considered as a sort of asymmetry between the dipolar and quadrupolar description of the body, so that questions may be raised on whether some physical consistency of the model would be necessary at this point.

A second general remark concerns the number of independent components of the relativistic quadrupole tensor $J$.
As stated above, this number is 20 and includes the mass quadrupole moment, the flow quadrupole moment and the stress quadrupole moment.
In the literature \cite{ehlers77} the mass quadrupole moment is also denoted by ${\mathcal M}(u)=Q_{\rm ER}(u)$, with
\beq
{\mathcal M}(u)^{\alpha\beta}=\frac43 J^{\alpha\gamma\beta\delta}u_\gamma  u_\delta=\frac43 Q( u)^{\alpha\beta }\,,
\eeq
and is used to form the moment of inertia of the body
\beq
I( u)^{\alpha\beta}={\mathcal M}( u)^{\gamma}{}_\gamma P(u)^{\alpha\beta}- {\mathcal M}( u)^{\alpha\beta}\,,
\eeq
$u$ being the (timelike) direction of the generalized momentum of the body.

Following Ehlers and Rudolph \cite{ehlers77}, it is possible to show that the relation between ${\mathcal M}( u)^{\alpha\beta}$ and $I( u)^{\alpha\beta}$
coincides with that existing between the corresponding familiar quantities of Newtonian mechanics. [Neverthless, one may also adopt different definitions of both mass quadrupole moment and moment of inertia of a body which equally reduce to their Newtonian counterparts in the Newtonian limits; so this is in only a possible choice, even if quite natural.]
Furthermore, through the spin vector and the moment of inertia it is possible to define an angular velocity vector for the body itself, namely
\beq
S( u)^\alpha= I( u)^{\alpha\beta}\Omega( u)_\beta\,, 
\eeq
with $\Omega( u)$ defined all along the center of mass line $U$ (and only there).
The introduction of $\Omega$ can be related in turn to a special spatial triad $\{e_{\hat a}\}$ ($a=1,2,3$) adapted to $u$, such that
\beq
P(u)^\alpha{}_\beta \nabla_U e_{\hat a}^\beta =\eta( u)^\alpha{}_{\mu\nu} \Omega( u)^\mu e_{\hat a}^\nu\,. 
\eeq
This triad is the relativistic analogue of the Newtonian \lq\lq body-fixed" spatial frame and it might be (at least formally and driven by the Newtonian analogy) convenient to describe the body dynamics in terms of the triad vectors, the angular velocity vector and the momentum of inertia.
Explicit evaluation of $\Omega(u)$ and the associated frame $e_{\hat a}$ can be done easily, but the resulting expressions are in general rather involved and of limited practical use (special situations only can be explored in this sense, e.g., $Q(u)$ diagonal in the proper frame or spin aligned with some preferred direction).

Finally, concerning the flow quadrupole moment $\pi ( u)^{\alpha\beta\gamma}$ and the stress quadrupole moment $\Pi (u)^{\alpha\beta\gamma}$, apart from the fact that they reduce to the corresponding Newtonian quantities in the Newtonian limit, there are neither relevant properties nor examples and specific literature. So far, their role in the relativistic extended body dynamics still necessitates further investigations.

\section{Dynamics of extended bodies in the equatorial plane of a Schwarzschild spacetime}

Let us begin by writing the Schwarzschild metric in the standard form
\beq
\label{sch}
ds^2=-N^2dt^2+N^{-2}dr^2+r^2(d\theta^2+\sin^2\theta d\phi^2)\,,
\eeq
with the \lq\lq lapse function" $N$ given by
\beq
N=\sqrt{1-\frac{2M}{r}}\,,
\eeq
($M$ is the mass of the central body). 
The metric (\ref{sch}) is static, i.e., $\partial_t$ is a hypersurface-forming timelike Killing vector. 
Observers at rest with respect to the coordinates (or \lq\lq static observers") have their $4$-velocity vector 
aligned along the Killing direction itself, namely
\beq
n=N^{-1}\partial_t\equiv e_{\hat t}\,.
\eeq
A natural orthonormal frame $e_{\hat a}$  ($a=1,2,3$) adapted to them is given by 
\begin{eqnarray}
\label{frame}
e_{\hat r}=N\partial_r\,,\quad e_{\hat \theta}=\frac{1}{r}\partial_\theta\,,\quad e_{\hat\phi}=\frac{1}{r\sin\theta}\partial_\phi\,,
\end{eqnarray}
with dual
\begin{eqnarray}
\omega^{\hat t}&=&N\rmd t=-n^\flat \,, \quad
\omega^{\hat r}=N^{-1}\rmd r\,, \nonumber\\
\omega^{\hat \theta}&=&r\rmd \theta\,, \quad
\omega^{\hat \phi}=r\sin\theta\rmd \phi\,,
\end{eqnarray}
$n^\flat$ being the fully covariant representation of $n$.

For a later use, let us recall the general timelike geodesic equations on the equatorial plane $\theta=\pi/2$
\begin{eqnarray}
\label{eqrtest}
\frac{\rmd t}{\rmd \tau} &=&\frac{E}{N}\,,\qquad
\frac{\rmd \phi}{\rmd \tau} =\frac{L}{r^2}\,,\nonumber\\
\frac{\rmd r}{\rmd \tau} &=&\pm\left[
E^2-N^2\left(1+\frac{L^2}{r^2}\right)
\right]^{1/2}\,,
\end{eqnarray}
where $E$ and $L$ are the conserved Killing energy and angular momentum per unit mass, respectively. 
Their explicit solutions can be expressed in terms of elliptic integrals, as it is well known.
In the special case of circular motion, co-rotating $(+)$ and counter-rotating $(-)$ circular geodesic orbits are characterized by 
the \lq\lq Keplerian" $4$-velocity
\beq
\label{Ugeocirc}
U_K=\gamma_K (n\pm \nu_K e_{\hat \phi})=\Gamma_K (\partial_t \pm \zeta_K \partial_\phi)\,,
\eeq
where $\gamma_K=(1-\nu^2_K)^{-1/2}$ with
\begin{eqnarray}
\label{kepler}
\Gamma_K&=&\frac{\gamma_K}{N}=\frac{\gamma_K\nu_K}{r\zeta_K}\,, \qquad
\nu_K=\sqrt{\frac{M}{r-2M}}\,,\nonumber\\
\gamma_K&=&\sqrt{\frac{r-2M}{r-3M}}\,, \qquad
\zeta_K=\sqrt{\frac{M}{r^3}}\,.
\end{eqnarray}

\subsection{Adapted frames, spin vector and quadrupole tensor}

Let the motion of the extended body be confined on the equatorial plane, i.e.,   
\beq
U=\gamma [n +\nu\hat \nu (U,n)]\,,
\eeq
with
\beq
\hat \nu (U,n)\equiv \hat \nu=\cos\alpha e_{\hat r}+ \sin\alpha e_{\hat \phi}\,,
\eeq
and $\gamma=(1-\nu^2)^{-1/2}$. 
A second orthonormal frame adapted to $n$ (besides $e_{\hat a}$) 
is built with the spatial triad
\begin{eqnarray}
\label{n_uframe}
E(n)_1&\equiv& \hat \nu ^\perp=\sin\alpha  e_{\hat r}- \cos\alpha  e_{\hat \phi}\,,\nonumber\\
E(n)_2&=&\hat \nu\,, \qquad
E(n)_3=-e_{\hat \theta}\,.
\end{eqnarray}
Since $U$ is obtained by boosting $n$ along $\hat \nu$, an adapted frame to $U$ is simply
\begin{eqnarray}
\label{U_frame}
E(U)_1&\equiv& \hat \nu ^\perp=\sin\alpha  e_{\hat r}- \cos\alpha  e_{\hat \phi}\,,\nonumber\\
E(U)_2&=&\gamma [\nu n+\hat \nu]\,,\quad
E(U)_3=-e_{\hat \theta}\,.
\end{eqnarray}  
Let us consider now the the 4-momentum $P=mu$, with
\beq
u=\gamma_u [n +\nu_u\hat \nu_u]
   \,, 
\eeq
and
\beq
\hat \nu (u,n)\equiv \hat \nu_u=\cos\alpha_u e_{\hat r}+ \sin\alpha_u e_{\hat \phi}\,,
\eeq
and $\gamma_u=(1-\nu_u^2)^{-1/2}$. 
Similarly, since $u$ is obtained by boosting $n$ along $\hat \nu_u$, an orthonormal frame adapted to $u\equiv e_0$ is then built with the spatial triad
\begin{eqnarray}
\label{uframe}
e_1&\equiv& \hat \nu_u^\perp=\sin\alpha_u e_{\hat r}- \cos\alpha_u e_{\hat \phi}\,,\nonumber\\
e_2&=&\gamma_u [\nu_u n +\hat \nu_u]\,,\quad
e_3=-e_{\hat \theta}=E_3(n)\,,
\end{eqnarray}
where $e_1$ (like $E_3(n)$) is orthogonal to both $n$ and $u$. Clearly, boosting back this frame onto the Local Rest Space of $n$, one  has immediately also a third orthonormal frame adapted to $n$
\begin{eqnarray}
\label{n_uframe2}
F(n)_1 &\equiv &\hat \nu_u ^\perp=\sin\alpha_u  e_{\hat r}- \cos\alpha_u  e_{\hat \phi}\,,\nonumber\\
F(n)_2 &=&\hat \nu_u\,,\quad
F(n)_3=-e_{\hat \theta}\,.
\end{eqnarray}

For a later use we evaluate the transport laws of $u$ and $e_a$ along $U$, i.e.,
\begin{eqnarray}
\nabla_Uu &=& \lambda_1e_1+\lambda_2e_2\,,\nonumber\\
\nabla_Ue_1 &=& \lambda_1u+\lambda_3e_2\,,\nonumber\\
\nabla_Ue_2 &=& \lambda_2u-\lambda_3e_1\,,\nonumber\\
\nabla_Ue_3 &=& 0\,,
\end{eqnarray}
where
\begin{eqnarray}
\lambda_1&=&\gamma_u\left[-\nu_u\left(\frac{\rmd \alpha_u}{\rmd \tau}+\frac{N}{r}\gamma\nu\sin\alpha\right)+\frac{M}{r^2N}\gamma\sin\alpha_u\right]\,,\nonumber\\
\lambda_2&=&\gamma_u^2\frac{\rmd \nu_u}{\rmd \tau}+\frac{M}{r^2N}\gamma\cos\alpha_u\,,\\
\lambda_3&=&\gamma_u\left[\left(\frac{\rmd \alpha_u}{\rmd \tau}+\frac{N}{r}\gamma\nu\sin\alpha\right)-\nu_u\frac{M}{r^2N}\gamma\sin\alpha_u\right]\,.\nonumber
\end{eqnarray}

The parallel transport of the vector $u$ (corresponding to the direction of generalized momentum)
along $U$ implies $\lambda_1=0=\lambda_2$, that is
\beq
u=U\,,
\eeq
i.e., $\nu_u=\nu$ and $\alpha_u=\alpha$ with $U$ geodesic so that 
\begin{eqnarray}
\label{geoeqs}
\frac{\rmd \alpha}{\rmd \tau} &=&
-\frac{N}{r}\gamma\nu\sin\alpha\left(1-\frac{M}{r N^2}\frac{1}{\nu^2}\right)\,,\nonumber\\
\frac{\rmd \nu}{\rmd \tau} &=& 
-\frac{M}{r^2N}\frac{1}{\gamma}\cos\alpha\,,
\end{eqnarray}
Taking then into account the evolution equations $U=\rmd x^\alpha/\rmd\tau$, i.e., 
\beq
\label{Ucompts}
\frac{\rmd t}{\rmd \tau} =\frac{\gamma}{N}\,,\quad
\frac{\rmd r}{\rmd \tau} =N\gamma\nu\cos\alpha\,,\quad
\frac{\rmd \phi}{\rmd \tau} =\frac{\gamma\nu}{r}\sin\alpha\,,
\eeq
allows to fully integrate Eq. (\ref{geoeqs}). 
In fact, by eliminating the dependence on the proper time in favor of $r$ through the second of the previous equations, the equation for $\nu$ becomes  
\beq
\frac{\nu }{1-\nu^2}\frac{\rmd \nu}{\rmd r} =
-\frac{M}{r^2N^2}  =-\frac{M}{r^2-2Mr}\,.
\eeq
Imposing the initial condition $\nu(r_0)=\nu_0$ (with the limit $\nu_0\to1$ for $r_0\to2M$) gives
\beq
\frac{1-\nu^2}{1-\nu_0^2}=\frac{r_0(r-2M)}{r(r_0-2M)}\,,
\eeq
i.e.,
\beq
N\gamma=N_0\gamma_0=E=const\,,
\eeq
so that 
\beq
\nu= \sqrt{1-\left(\frac{N}{N_0\gamma_0}\right)^2}\,.
\eeq
Similarly, the equation for $\alpha$ as a function of $r$ turns out to be 
\beq
\frac{1}{ \tan\alpha}\frac{\rmd \alpha}{\rmd r}=-\frac{1}{r}\left(1-\frac{M}{r N^2 \nu^2}\right)\,, 
\eeq
whose solution is given by
\beq
r\gamma \nu\sin\alpha=r_0\gamma_0 \nu_0\sin\alpha_0=L=const\,.
\eeq
The evolution equations (\ref{Ucompts}) giving the dependence of the coordinates on the proper time can then be written in terms of the conserved Killing quantities $E$ and $L$ as in Eq. (\ref{eqrtest}). 
The above conditions, i.e., $U=u$ geodesic and $e_3$ parallely transported along $U$, fully determine the equatorial motion of a structureless particle. 
Let us see how these conditions modify due to the dipolar and quadrupolar structure of the particle.

The projection of the spin tensor into the local rest space of $u$ defines the spin vector $S(u)$ (hereafter simply denoted by $S$, for short).
When decomposed with respect to the frame adapted to $n$ the spin vector is then given by
\beq
S=S^{\hat t}e_{\hat t}+S^{\hat r}e_{\hat r}+S^{\hat \theta}e_{\hat \theta}+S^{\hat \phi}e_{\hat \phi}\,,
\eeq
with $S^{\hat t}=\nu_u[S^{\hat r}\cos\alpha_u+S^{\hat \phi}\sin\alpha_u]$ due to the supplementary conditions (\ref{tulczconds}).
When decomposed with respect to the frame (\ref{uframe}) adapted to $u$ it writes instead as
\beq
S=S^1e_1+S^2e_2+S^3e_3\,.
\eeq

Finally, let the quadrupole tensor be given by Eq. (\ref{deco_bar_u2}) in terms of its electric, magnetic and mixed parts (\ref{Jsplit_baru}) with $\bar u=u$.
When expressed with respect to $n$, $J$ has the following nonvanishing 21 frame components
\begin{eqnarray}
&(6)\quad&
J_{\hat t \hat r \hat t \hat r}=Q(n)_{\hat r \hat r}\,,\quad
J_{\hat t \hat \theta \hat t \hat \theta}=Q(n)_{\hat \theta \hat \theta}\,,\nonumber\\
&&
J_{\hat t \hat \phi \hat t \hat \phi}=Q(n)_{\hat \phi \hat \phi}\,,\quad
J_{\hat t \hat r \hat t \hat \theta}=Q(n)_{\hat r \hat \theta}\,,\nonumber\\
&& 
J_{\hat t \hat r \hat t \hat \phi}=Q(n)_{\hat r \hat \phi}\,,\quad
J_{\hat t \hat \theta \hat t \hat \phi}=Q(n)_{\hat \theta \hat \phi}\,,\nonumber\\
&(9)\quad&
J_{\hat t \hat r \hat \theta \hat \phi}=-W(n)_{\hat r \hat r}\,,\quad
J_{\hat t \hat \theta \hat \theta \hat \phi}=-W(n)_{\hat \theta\hat r }\,,\nonumber\\ 
&&  
J_{\hat t \hat \phi \hat \theta \hat \phi}=-W(n)_{\hat \phi\hat r }\,,\quad 
J_{\hat t \hat r \hat r \hat \phi}=W(n)_{\hat r \hat \theta }\,,\nonumber\\ 
&&  
J_{\hat t \hat \theta \hat r \hat \phi}=W(n)_{\hat \theta \hat \theta }\,,\quad
J_{\hat t \hat \phi \hat r \hat \phi}=W(n)_{\hat \phi \hat \theta }\,,\nonumber\\ 
&&
J_{\hat t \hat r \hat r \hat \theta}=-W(n)_{\hat r\hat \phi }\,,\quad 
J_{\hat t \hat \theta \hat r \hat \theta}=-W(n)_{\hat \theta\hat \phi }\,,\nonumber\\ 
&& 
J_{\hat t \hat \phi \hat r \hat \theta}=-W(n)_{\hat \phi \hat \phi}\,,\nonumber\\
&(6)\quad&
J_{\hat \theta \hat \phi \hat \theta \hat \phi}=M(n)_{\hat r \hat r}\,,\quad
J_{\hat r \hat \phi \hat r \hat \phi}=M(n)_{\hat \theta \hat \theta}\,,\nonumber\\
&&
J_{\hat r \hat \theta \hat r \hat \theta}=M(n)_{\hat \phi \hat \phi}\,,\quad
J_{\hat r \hat \phi \hat \theta \hat \phi}=-M(n)_{\hat r \hat \theta}\,,\nonumber\\
&&
J_{\hat r \hat \theta \hat \theta \hat \phi}=M(n)_{\hat r \hat \phi}\,,\quad
J_{\hat r \hat \theta \hat r \hat \phi}=-M(n)_{\hat \theta \hat \phi}\,.
\end{eqnarray}
The additional constraint $J_{[\alpha\beta\gamma]\delta}=0$ implies that the magnetic part be trace-free, i.e.,
\beq
0=W(n)_{\hat r \hat r}+W(n)_{\hat \theta \hat \theta }+W(n)_{\hat \phi \hat \phi}
\,,
\eeq
so that the independent components of $J$ reduce to 20.

\subsection{Spin terms}

Consider first the contribution to the force term due to spin, i.e.,
\beq
\label{fspindef}
F^\mu_{\rm (spin)}=
- \frac12 \, R^\mu{}_{\nu \alpha \beta} \, U^\nu \, S^{\alpha \beta}\,.
\eeq 
When decomposed with respect to the frame adapted to $n$ it becomes
\begin{widetext}
\begin{eqnarray}
F_{\rm (spin)}&=&-\frac{M}{r^3}\gamma\gamma_u\left\{
\nu\nu_u(2\sin\alpha_u\cos\alpha+\sin\alpha\cos\alpha_u)S^{\hat \theta}n
+(2\nu_u\sin\alpha_u+\nu\sin\alpha)S^{\hat \theta}e_{\hat r}\right.\nonumber\\
&&\left.
+\{[2\nu\sin\alpha(1-\nu_u^2\cos^2\alpha_u)+\nu_u\sin\alpha_u(1-\nu\nu_u\cos\alpha\cos\alpha_u)]S^{\hat r}\right.\nonumber\\
&&\left.
+[\nu\cos\alpha(1-\nu_u^2\sin^2\alpha_u)-\nu_u\cos\alpha_u(1+2\nu\nu_u\sin\alpha\sin\alpha_u)]S^{\hat \phi}
\}e_{\hat \theta}\right.\nonumber\\
&&\left.
+(\nu_u\cos\alpha_u-\nu\cos\alpha)S^{\hat \theta}e_{\hat \phi}
\right\}\,,
\end{eqnarray}
whereas when decomposed with respect to the frame adapted to $u$ it writes
\begin{eqnarray}
F_{\rm (spin)}&=&\frac{M}{r^3}\gamma\gamma_u\left\{
3\gamma_u\nu_u\sin\alpha_u(\nu\cos\alpha-\nu_u\cos\alpha_u)S^3u
+\frac12[\nu_u(1-3\cos2\alpha_u)+2\nu\cos(\alpha_u-\alpha)]S^3e_1\right.\nonumber\\
&&\left.
-\frac12\gamma_u[(2+\nu_u^2)\nu\sin(\alpha_u-\alpha)+3\nu_u(\nu\nu_u\sin(\alpha_u+\alpha)-\sin2\alpha_u)]S^3e_2\right.\nonumber\\
&&\left.
+\left[
[\nu_u+\frac12\nu(\cos(\alpha_u-\alpha)-3\cos(\alpha_u+\alpha))]S^1
+\frac12\frac{\nu}{\gamma_u}[3\sin(\alpha_u+\alpha)-\sin(\alpha_u-\alpha)]S^2
\right]e_3
\right\}\,.\nonumber\\
\end{eqnarray}
\end{widetext}
Finally
\begin{eqnarray}
D_{\rm (spin)}&=&m\gamma\gamma_u\left[
(\nu_u\cos\alpha_u-\nu\cos\alpha)\omega^{\hat t}\wedge\omega^{\hat r}\right.\nonumber\\
&&\left.
+(\nu_u\sin\alpha_u-\nu\sin\alpha)\omega^{\hat t}\wedge\omega^{\hat \phi}\right.\nonumber\\
&&\left.
-\nu_u\nu\sin(\alpha_u-\alpha)\omega^{\hat r}\wedge\omega^{\hat \phi}
\right]\nonumber\\
&=&-\omega^{\hat t}\wedge{\mathcal E}(n)_{\rm (spin)}+{}^{*_{(n)}}{\mathcal B}(n)_{\rm (spin)}\,,
\end{eqnarray}
which identifies 
\begin{eqnarray}
{\mathcal E}(n)_{\rm (spin)}&=&-m\gamma\gamma_u\left[
(\nu_u\cos\alpha_u-\nu\cos\alpha)\omega^{\hat r}\right.\nonumber\\
&&\left.
+(\nu_u\sin\alpha_u-\nu\sin\alpha)\omega^{\hat \phi}\right]\,,\nonumber\\
{\mathcal B}(n)_{\rm (spin)}&=&m\gamma\gamma_u\nu_u\nu\sin(\alpha_u-\alpha)\omega^{\hat \theta}\,,
\end{eqnarray}
with ${\mathcal E}(n)_{\rm (spin)}\cdot{\mathcal B}(n)_{\rm (spin)}=0$.
Similarly
\begin{eqnarray}
D_{\rm (spin)}&=&-m\gamma\left[
\nu\sin(\alpha_u-\alpha)\omega^{0}\wedge\omega^{1}\right.\nonumber\\
&&\left.
+\gamma_u(\nu\cos(\alpha_u-\alpha)-\nu_u)\omega^{0}\wedge\omega^{2}
\right]\nonumber\\
&=&-\omega^0\wedge\left\{m\gamma[\nu\sin(\alpha_u-\alpha)\omega^{1}\right.\nonumber\\
&&\left.
+\gamma_u(\nu\cos(\alpha_u-\alpha)-\nu_u)\omega^{2}]\right\}\nonumber\\
&\equiv&-\omega^0\wedge{\mathcal E}(u)_{\rm (spin)}\,,
\end{eqnarray}
with
\beq
{\mathcal E}(u)_{\rm (spin)}^2={\mathcal E}(n)_{\rm (spin)}^2-{\mathcal B}(n)_{\rm (spin)}^2\,.
\eeq

\subsection{Quadrupole terms}

Consider then the quadrupole contribution to the force term, i.e., 
\beq
\label{fquaddef}
F^\mu_{\rm (quad)}=
-\frac16 \, \, J^{\alpha \beta \gamma \delta} \,\nabla^\mu  R_{\alpha \beta \gamma \delta}\,.
\eeq 
With respect to the frame adapted to $n$ it becomes
\begin{eqnarray}
\label{Fquadzamo}
F_{\rm (quad)}&=&-4N\frac{M}{r^4}\bigg\{
\frac14[-6X(n)_{\hat r \hat r}+K]e_{\hat r}\nonumber\\
&&+X(n)_{\hat r \hat \theta}e_{\hat \theta}
+X(n)_{\hat r \hat \phi}e_{\hat \phi}
\bigg\}\,,
\end{eqnarray} 
where
\beq
K=-2(X(n)_{\hat r \hat r}+X(n)_{\hat \theta \hat \theta}+X(n)_{\hat \phi \hat \phi})\,,
\eeq
and
\beq
X(n)_{\hat a\hat b}=M(n)_{\hat a\hat b}-Q(n)_{\hat a\hat b}\,.
\eeq

Finally, the torque term, i.e., 
\beq
D^{\mu \nu}_{\rm (quad)}=\frac43 \, J^{\alpha \beta \gamma [\mu}R^{\nu]}{}_{\gamma \alpha \beta}\,,
\eeq
turns out to be given by
\begin{eqnarray}
D_{\rm (quad)}&=&4\frac{M}{r^3}\bigg[
2W(n)_{(\hat r \hat \phi)}\omega^{\hat t}\wedge\omega^{\hat \theta}
-2W(n)_{(\hat r \hat \theta)}\omega^{\hat t}\wedge\omega^{\hat \phi}\nonumber\\
&&+X(n)_{\hat r \hat \theta}\omega^{\hat r}\wedge\omega^{\hat \theta}
+X(n)_{\hat r \hat \phi}\omega^{\hat r}\wedge\omega^{\hat \phi}
\bigg]\nonumber\\
&=&-\omega^{\hat t}\wedge{\mathcal E}(n)_{\rm (quad)}+{}^{*_{(n)}}{\mathcal B}(n)_{\rm (quad)}\,,
\end{eqnarray}
which identifies 
\begin{eqnarray}
{\mathcal E}(n)_{\rm (quad)}&=&-\frac{8M}{r^3}\left[
W(n)_{(\hat r \hat \phi)}\omega^{\hat \theta}-W(n)_{(\hat r \hat \theta)}\omega^{\hat \phi}
\right]\,,\nonumber\\
{\mathcal B}(n)_{\rm (quad)}&=&4\frac{M}{r^3}\left[X(n)_{\hat r \hat \theta}\omega^{\hat \phi}-X(n)_{\hat r \hat \phi}\omega^{\hat \theta}\right]\,.
\end{eqnarray}
Note that the quadrupole force can be also written as
\begin{eqnarray}
\label{Fquadzamo2}
F_{\rm (quad)}&=&-4N\frac{M}{r^4}\bigg\{
\frac14[-6X(n)_{\hat r \hat r}+K]e_{\hat r}\nonumber\\
&&+e_{\hat r}\times {\mathcal B}(n)_{\rm (quad)}
\bigg\}\,.
\end{eqnarray}

Since we have shown that there exist various models and approaches (i.e., an arbitrariness often ascribed to physical reasons or mathematical simplifications) in expressing the quadrupolar structure of the body, we will consider below specific cases associated with different choices.

\subsection{MPD set of equations}

Let us consider the special case in which the spin vector is aligned along the $z$-axis, i.e.,
\beq
S=se_3\,.
\eeq
The spin force (with the property $F_{\rm (spin)}\cdot U=0$) becomes
\begin{widetext} 
\begin{eqnarray}
F_{\rm (spin)}&=&\frac{M}{r^3}\gamma\gamma_u s\left\{
3\gamma_u\nu_u\sin\alpha_u(\nu\cos\alpha-\nu_u\cos\alpha_u)u
+\frac12[\nu_u(1-3\cos2\alpha_u)+2\nu\cos(\alpha_u-\alpha)]e_1\right.\nonumber\\
&&\left.
-\frac12\gamma_u[(2+\nu_u^2)\nu\sin(\alpha_u-\alpha)+3\nu_u(\nu\nu_u\sin(\alpha_u+\alpha)-\sin2\alpha_u)]e_2
\right\}\,.
\end{eqnarray}
\end{widetext}

Consider then the quadrupole force given by Eq. (\ref{Fquadzamo}) with respect to the frame adapted to $n$. 
In order to write the corresponding expression in the $u$-frame we need both to transform the frame (according to Eq. (\ref{uframe})) and to express the quantities $X(n)_{\hat a\hat b}$ in terms of the $X(u)_{ab}$.
To simplify the description of the extended body one can assume that the quadrupole tensor in the $u$-frame be represented by two independent components only, i.e., $X(u)_{11}$ and $X(u)_{22}$, with $X(u)_{33}=-X(u)_{11}-X(u)_{22}$, the remaining components being set equal to zero (namely, $W(u)_{ab}=0$ and also all the nondiagonal components of $X(u)_{ab}$).
Imposing such a condition can be easily relaxed and has only the result of make the formulas more compact, without affecting the main features of the underlying physics.
Using then the following transformation laws 
\begin{eqnarray}
X(n)_{\hat r\hat r}&=&\gamma_u^2\{X(u)_{11}-\cos^2\alpha_u[(X(u)_{11}+2X(u)_{22})\nu_u^2\nonumber\\
&&+X(u)_{11}-X(u)_{22}]
+\nu_u^2(X(u)_{11}+X(u)_{22})\}\,,\nonumber\\
X(n)_{\hat r\hat \theta}&=&0\,,\nonumber\\
X(n)_{\hat r\hat \phi}&=&-\gamma_u^2\sin\alpha_u\cos\alpha_u[(X(u)_{11}+2X(u)_{22})\nu_u^2\nonumber\\
&&+X(u)_{11}-X(u)_{22}]\,,\nonumber\\
X(n)_{\hat \theta\hat \theta}&=&-\gamma_u^2\left[X(u)_{11}\nu_u^2+X(u)_{11}+X(u)_{22}\right]\,,\nonumber\\
X(n)_{\hat \theta\hat \phi}&=&0\,,\nonumber\\
X(n)_{\hat \phi\hat \phi}&=&-X(n)_{\hat r\hat r}-X(n)_{\hat \theta\hat \theta}\,,
\end{eqnarray}
the quadrupole force becomes
\begin{eqnarray}
\label{Fquadframeu}
F_{\rm (quad)}&=&F_{\rm (quad)}^1e_1+F_{\rm (quad)}^2(-\nu_u u+e_2)\nonumber\\
&=&F_{\rm (quad)}^1e_1+F_{\rm (quad)}^2\frac{\hat \nu_u}{\gamma_u}\,,
\end{eqnarray}
with 
\begin{widetext}
\begin{eqnarray}
F_{\rm (quad)}^1&=&
2N\frac{M}{r^4}\gamma_u^2\sin\alpha_u\left\{
5\cos2\alpha_u[X(u)_{22}(1-2\nu_u^2)-X(u)_{11}(1+\nu_u^2)]+X(u)_{22}(1+4\nu_u^2)+5X(u)_{11}(1+\nu_u^2)
\right\}
\,,\nonumber\\
F_{\rm (quad)}^2&=&
2N\frac{M}{r^4}\gamma_u^3\cos\alpha_u\left\{
5\cos2\alpha_u[X(u)_{22}(1-2\nu_u^2)-X(u)_{11}(1+\nu_u^2)]+X(u)_{22}(5-4\nu_u^2)+X(u)_{11}(1+\nu_u^2)
\right\}
\,,\nonumber\\
\end{eqnarray}
\end{widetext}
and $\gamma_u(-\nu_u u+e_2)$ represents a unitary and spacelike vector orthogonal to $n$.
Similarly, the torque term turns out to be
\begin{eqnarray}
D_{\rm (quad)}&=&-\frac{2M}{r^3}\gamma_u\left[
\nu_u\sin2\alpha_u(X(u)_{11}+2X(u)_{22})\omega^{0}\wedge\omega^{1}\right.\nonumber\\
&&\left.
-2\gamma_u\nu_u\sin^2\alpha_u(2X(u)_{11}+X(u)_{22})\omega^{0}\wedge\omega^{2}\right.\nonumber\\
&&\left.
+\sin2\alpha_u(X(u)_{11}-X(u)_{22})\omega^{1}\wedge\omega^{2}
\right]\nonumber\\
&=&-\omega^0\wedge{\mathcal E}(u)_{\rm (quad)}+{}^{*_{(u)}}{\mathcal B}(u)_{\rm (quad)}\,.
\end{eqnarray}
The latter equation identifies 
\begin{eqnarray}
{\mathcal E}(u)_{\rm (quad)}&=&\frac{2M}{r^3}\gamma_u\nu_u\left[
\sin2\alpha_u(X(u)_{11}+2X(u)_{22})\omega^{1}\right.\nonumber\\
&&\left.
-2\gamma_u\sin^2\alpha_u(2X(u)_{11}+X(u)_{22})\omega^{2}\right]\,,\nonumber\\
{\mathcal B}(u)_{\rm (quad)}&=&-\frac{2M}{r^3}\gamma_u\sin2\alpha_u(X(u)_{11}-X(u)_{22})\omega^{3}\,,\nonumber\\
\end{eqnarray}
with ${\mathcal E}(u)_{\rm (quad)}\cdot{\mathcal B}(u)_{\rm (quad)}=0$.
Note that with this choice of nonzero components of the quadrupole tensor we have also
\beq
{\mathcal E}(n)_{\rm (quad)}=0\,,\quad
{\mathcal B}(n)_{\rm (quad)}=-4\frac{M}{r^3}X(n)_{\hat r \hat \phi}\omega^{\hat \theta}\,.
\eeq

The whole set of MPD equations then reduces to 
\begin{eqnarray} 
\label{setfin}
\frac{\rmd m}{\rmd \tau} &=& 
F_{\rm (spin)}^0-\nu_u F_{\rm (quad)}^2\,,\nonumber\\
\frac{\rmd \alpha_u}{\rmd \tau} &=& 
\frac{N}{r}\gamma\nu\sin\alpha 
-\frac{M}{r^2N}\frac{\gamma}{\nu_u}\sin\alpha_u\nonumber\\
&&-\frac{1}{m\gamma_u\nu_u}(F_{\rm (spin)}^1 + F_{\rm (quad)}^1)\,,\nonumber\\
\frac{\rmd \nu_u}{\rmd \tau} &=& 
-\frac{M}{r^2N}\frac{\gamma}{\gamma_u^2}\cos\alpha_u
+\frac{1}{m\gamma_u^2}(F_{\rm (spin)}^2 + F_{\rm (quad)}^2)\,,\nonumber\\
\frac{\rmd s}{\rmd \tau} &=&
{\mathcal B}(u)_{\rm (quad)}^3= 
-\frac{2M}{r^3}\gamma_u\sin2\alpha_u(X(u)_{11}-X(u)_{22})\,,\nonumber\\
\end{eqnarray}
together with the following two compatibility conditions coming from the spin evolution equations 
\begin{eqnarray} 
0&=&m^2\gamma\nu\sin(\alpha_u-\alpha)-mD_{\rm (quad)}{}_{01}\nonumber\\
&&+s(F_{\rm (spin)}^2 + F_{\rm (quad)}^2)\,,\nonumber\\
0&=&m^2\gamma\gamma_u[\nu_u-\nu\cos(\alpha_u-\alpha)]+mD_{\rm (quad)}{}_{02}\nonumber\\
&&+s(F_{\rm (spin)}^1 + F_{\rm (quad)}^1)\,,
\end{eqnarray}
which give two algebraic relations involving the remaining unknowns $\nu$ and $\alpha$.
After some manipulation we find
\beq
\label{nuandalpha}
\tan\alpha=\frac{A+B\gamma}{C+D\gamma}\,, \qquad
\gamma=\frac{k_1+\sqrt{k_1^2+k_2k_3}}{k_2}\,,
\eeq
where
\begin{eqnarray} 
k_1&=&AB+CD\,,\nonumber\\
k_2&=&1-B^2-D^2\,,\nonumber\\
k_3&=&1+A^2+C^2\,,
\end{eqnarray}
and 
\begin{eqnarray} 
\label{ABC_etc}
A&=&\frac{1}{\tilde A}\left[
\xi_1\sin\alpha_u\left(sF_{\rm (quad)}^1
-m{\mathcal E}(u)_{\rm (quad)}^2
\right)\right.\nonumber\\
&&\left.
+\xi_2\cos\alpha_u
\left(sF_{\rm (quad)}^2
+m{\mathcal E}(u)_{\rm (quad)}^1
\right)
\right]\,,\nonumber\\
B&=&\nu_u\sin\alpha_u\frac{m^2+2s^2\zeta_K^2}{m^2-s^2\zeta_K^2}\,,\nonumber\\
C&=&\frac{\xi_2}{\gamma_u\tilde A}\left[
\cos\alpha_u\left(sF_{\rm (quad)}^1
-m{\mathcal E}(u)_{\rm (quad)}^2
\right)\right.\nonumber\\
&&\left.
-\gamma_u\sin\alpha_u\left(sF_{\rm (quad)}^2
+m{\mathcal E}(u)_{\rm (quad)}^1
\right)
\right]\,,\nonumber\\
D&=&\nu_u\cos\alpha_u\,,
\end{eqnarray}
with $D_{\rm (quad)}{}_{0a}=-{\mathcal E}(u)_{\rm (quad)}^a$ and
\begin{eqnarray}
{\tilde A}&=&-\gamma_u\xi_2(\xi_2\cos^2\alpha_u+\gamma_u\xi_1\sin^2\alpha_u)\,,\nonumber\\
\xi_1&=&(1+2\nu_u^2)s^2\zeta_K^2-\frac{m^2}{\gamma_u^2}\,,\nonumber\\
\xi_2&=&\frac{s^2\zeta_K^2-m^2}{\gamma_u}\,.
\end{eqnarray}
Note that the following relations hold
\beq
\nu\sin\alpha=\frac{A}{\gamma}+B\,,\qquad
\nu\cos\alpha=\frac{C}{\gamma}+D\,,
\eeq
whence $\nu\cos\alpha-\nu_u\cos\alpha_u=C/\gamma$, implying that the first equation of Eqs. (\ref{setfin}) governing the mass evolution becomes
\beq
\frac{\rmd m}{\rmd \tau} = \nu_u\left(3\zeta_K^2\gamma_u^2\sin\alpha_u sC- F_{\rm (quad)}^2\right)\,.
\eeq
The rhs of this equation vanishes for vanishing quadrupole (i.e., $F_{\rm (quad)}^a=0=D_{\rm (quad)}{}_{ab}$ implying $C=0$), yielding the well known constant mass result for a purely spinning particle.

Finally, the evolution equations (\ref{Ucompts}) must be also taken into account.

\begin{figure*} 
\typeout{*** EPS figure 1}
\begin{center}
$\begin{array}{cc}
\includegraphics[scale=0.4]{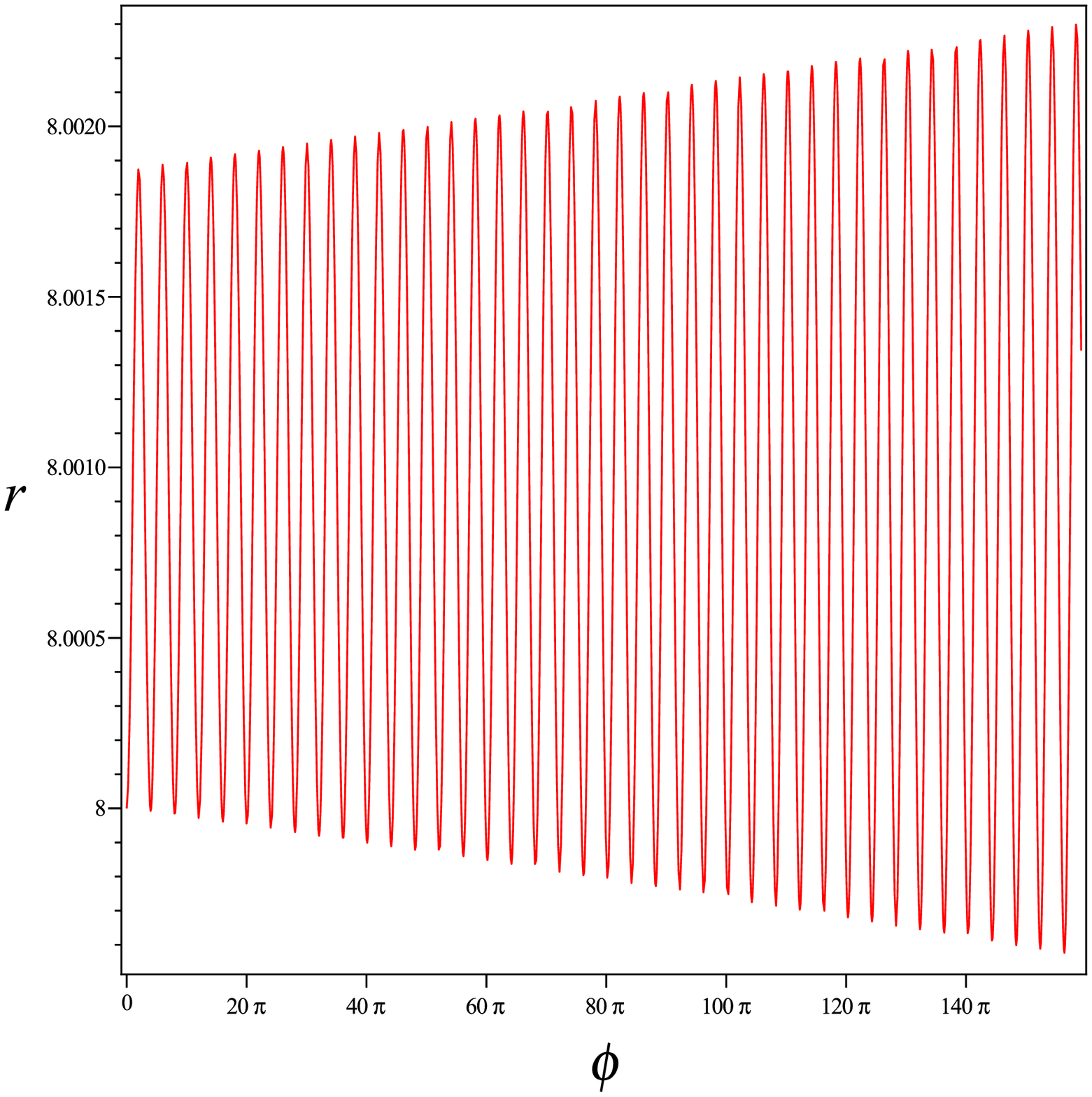}&\quad
\includegraphics[scale=0.4]{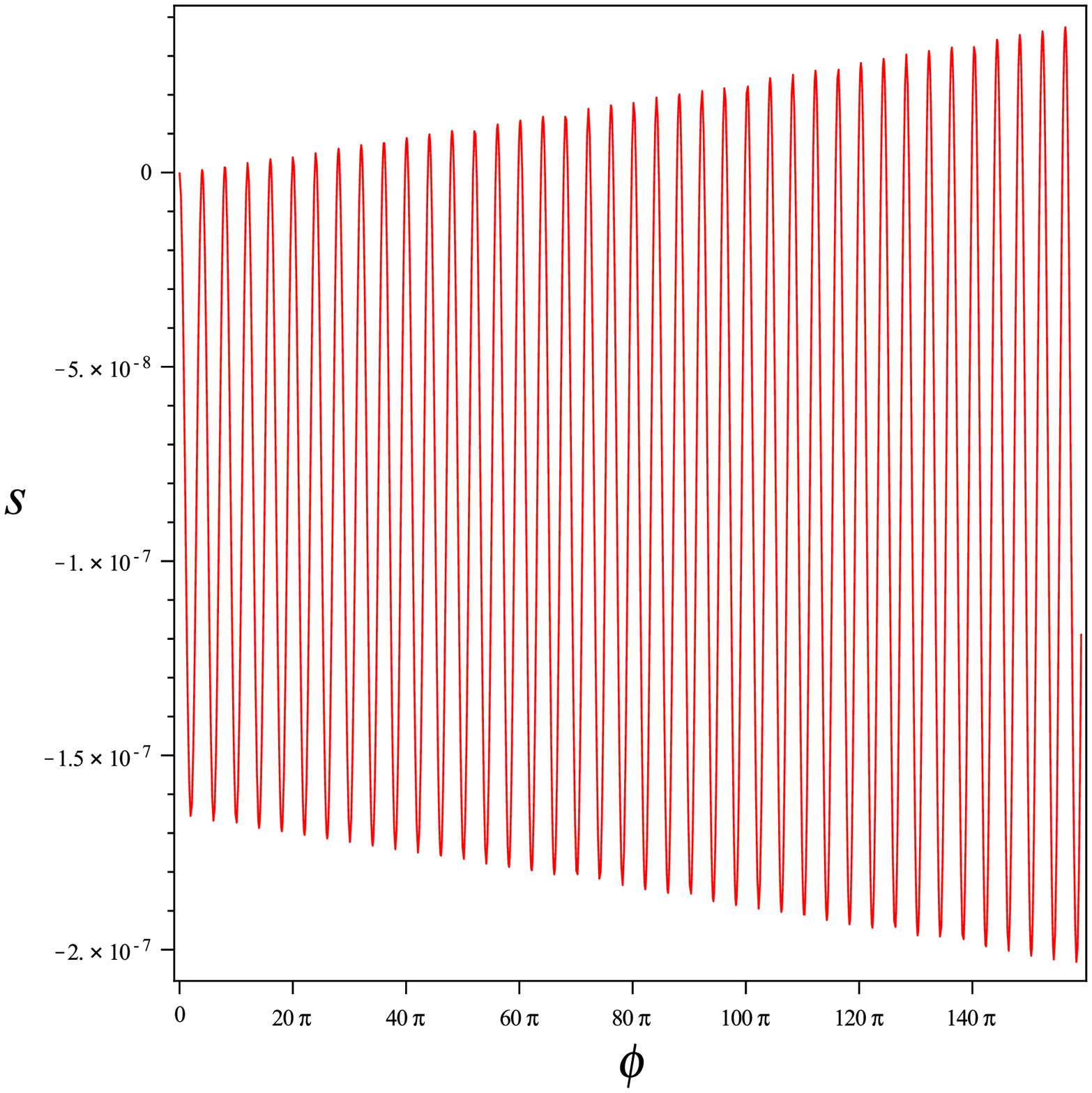}\\[.4cm]
\quad\mbox{(a)}\quad &\quad \mbox{(b)}
\end{array}$\\
\end{center}
\caption{The behavior of the radial coordinate is shown in panel (a) as a function of the azimuthal coordinate by solving the whole set of MPD equations numerically with the following choice of parameters and initial conditions: $r_0/M=8$, $X(u)_{11}/(m_0M^2)=0.001$, $X(u)_{22}/(m_0M^2)=-0.003$ and $r(0)=r_0$, $\phi(0)=0$, $\alpha_u(0)=\pi/2$, $\nu_u(0)=\nu_K\approx0.041$, $m(0)=m_0$, $s(0)=0$.
Panel (b) shows instead the corresponding behavior of the spin invariant.
Note that $r$ is expressed in units of $M$, whereas $s$ in units of both $m_0$ and $M$.
}
\label{fig:1}
\end{figure*}

\begin{figure*} 
\typeout{*** EPS figure 2}
\begin{center}
$\begin{array}{cc}
\includegraphics[scale=0.4]{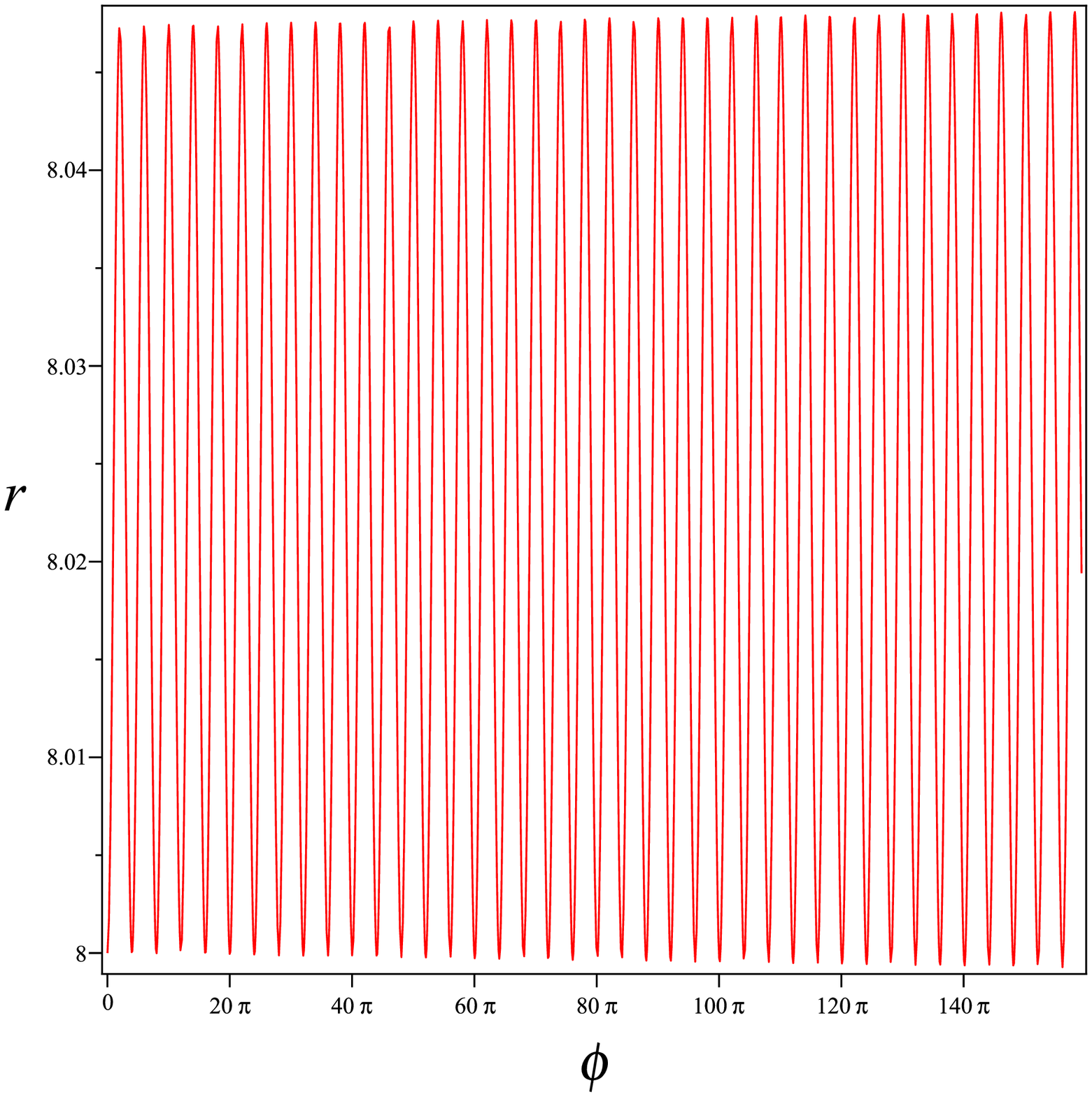}&\quad
\includegraphics[scale=0.4]{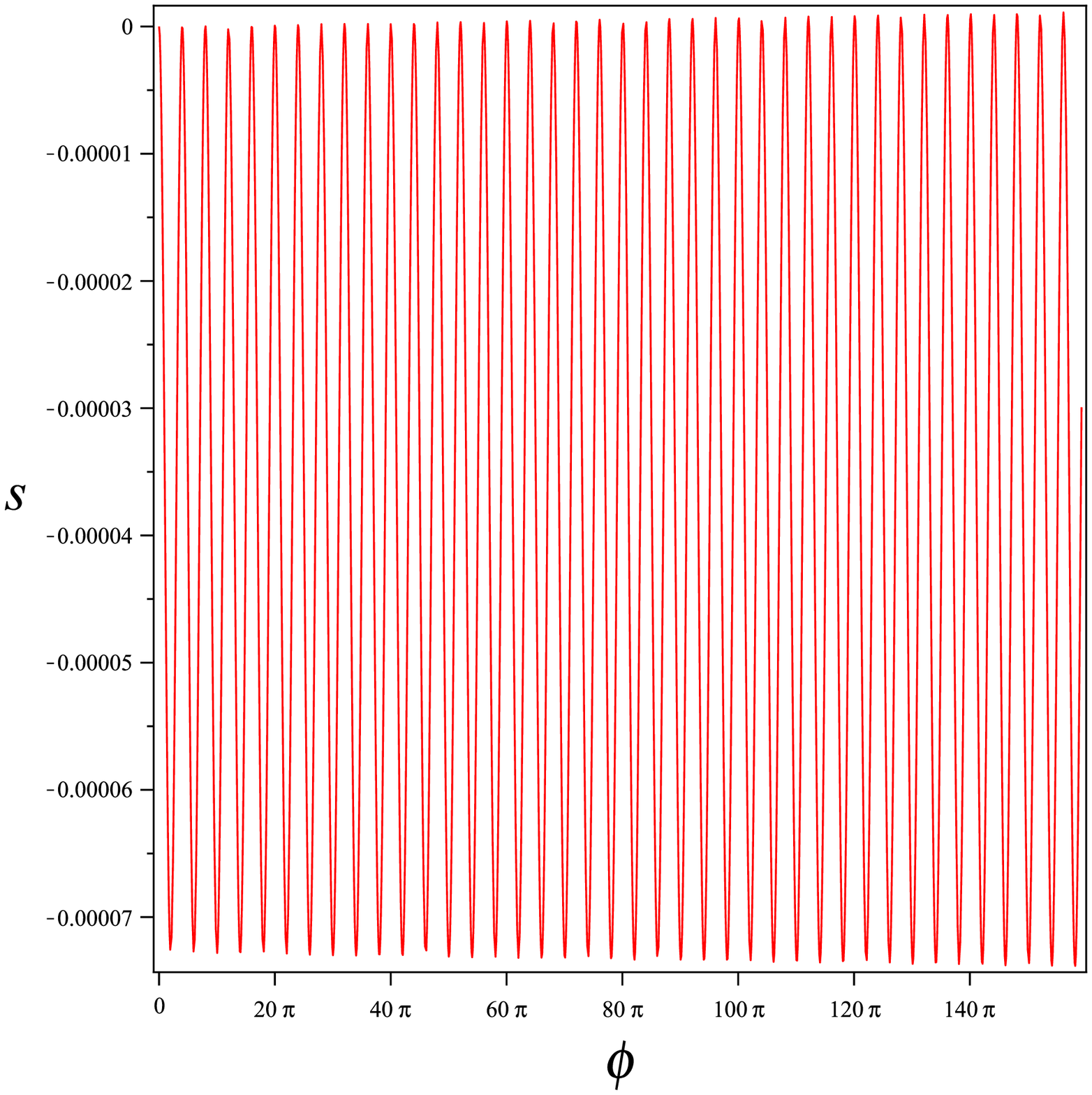}\\[.4cm]
\quad\mbox{(a)}\quad &\quad \mbox{(b)}
\end{array}$\\
\end{center}
\caption{The behaviors of the radial coordinate and of the spin invariant as functions of the azimuthal coordinate are shown in panels (a) and (b) respectively by solving the whole set of MPD equations numerically with the same initial conditions as in Fig. \ref{fig:1}, but for a different choice of the quadrupole parameters, i.e., $X(u)_{11}/(m_0M^2)=0.02$ and $X(u)_{22}/(m_0M^2)=-0.05$.
}
\label{fig:2}
\end{figure*}

Two examples of numerical integration of the MPD equations are shown in Figs. \ref{fig:1} and \ref{fig:2}.
The initial conditions have been chosen in both cases so that the $U$-trajectory is initially aligned with a stable  equatorial circular geodesic at a given value of the radial coordinate and the body is initially not spinning.
Figs. \ref{fig:1} (a) and (b) show the behaviors of the radial coordinate and of the spin invariant respectively as functions of the azimuthal coordinate in the case of very small values of the nonvanishing constant frame components $X(u)_{11}$ and $X(u)_{22}$ of the quadrupole tensor.
We see that the motion is confined inside a band close to the circular geodesic whose thickness slightly increases after each revolution.
This feature is definitely new with respect to the case of a purely spinning particle, where the orbit oscillates filling a circular corona of fixed width \cite{mashsingh,spin_dev_schw}. 
The occurrence of such a secular increase of the bandwidth for small values of the quadrupole actually characterizes the behaviors of all quantities involved and is confirmed by the approximated solution presented in the next section.
Obviously, this is a higher order effect only.
In fact, if the body is initially endowed with spin the oscillation amplitude is almost fixed by the value of the spin parameter. 
Furthermore, when the quadrupole parameters are not so small, this feature disappears as well, as shown in Fig. \ref{fig:2}, where the values of $X(u)_{11}$ and $X(u)_{22}$ are about 10 times greater than before.  

The quadrupolar structure of the body is responsible for the onset of spin angular momentum.
This is a direct consequence of the evolution equation for the spin invariant (see the last equation of  (\ref{setfin})), whose behavior as a function of the azimuthal coordinate is shown in Figs. \ref{fig:1} (b) and \ref{fig:2} (b).


\subsection{The case of quadrupole tensor proportional to the Riemann tensor}

Let us consider the case in which the quadrupole tensor proportional to the Riemann tensor as in Eq. (\ref{JproptoR}).
The quadrupole force has the same form as in Eq. (\ref{Fquadframeu}) with 
\begin{widetext}
\begin{eqnarray}
F_{\rm (quad)}^1&=&120\frac{N}{r}\zeta_K^4\gamma_u^4\sin\alpha_u\left\{
(c_1-c_2)\nu_u^2\cos^2\alpha_u\left[\nu_u^2\cos^2\alpha_u-\frac45(1+\nu_u^2)\right]
+\frac15[c_1(1+\nu_u^4)+(c_1-3c_2)\nu_u^2]
\right\}\,,\nonumber\\
F_{\rm (quad)}^2&=&120\frac{N}{r}\zeta_K^4\gamma_u^5\cos\alpha_u\left\{
(c_1-c_2)\nu_u^2\cos^2\alpha_u\left[\nu_u^2\cos^2\alpha_u-\frac25(2+3\nu_u^2)\right]\right.\nonumber\\
&&\left.
+\frac15[c_1+(2c_1-c_2)\nu_u^4+2(c_1-2c_2)\nu_u^2]
\right\}\,.
\end{eqnarray}
\end{widetext}
The torque term instead turns out to be given by
\beq
\label{torqueriem}
D_{\rm (quad)}=-\omega^0\wedge{\mathcal E}(u)_{\rm (quad)}\,,
\eeq
where 
\begin{eqnarray} 
{\mathcal E}(u)_{\rm (quad)}&=&
24(c_1-c_2)\zeta_K^4\gamma_u^3\nu_u\sin2\alpha_u\bigg[\frac12(1+\nu_u^2)\nonumber\\
&&-\nu_u^2\cos^2\alpha_u\bigg](\omega^1-\gamma_u\tan\alpha_u\omega^2)\,.
\end{eqnarray}
The evolution equations for $m$, $\nu_u$ and $\alpha_u$ have the same form as in Eq. (\ref{setfin}), whereas the spin invariant turns out to be constant because ${\mathcal B}(u)_{\rm (quad)}\equiv0$ in this case.
The numerical integration of the equations of motion with the same initial conditions as before, i.e., with the center of mass line initially aligned with a stable equatorial circular geodesic, gives again an oscillatory behavior of the radial coordinate.
If the body is not spinning, we find the same feature already shown in Figs. \ref{fig:1} (a) and \ref{fig:2} (a), i.e., the oscillations are confined in a region whose thickness slightly increases after each revolution if the values of the constants $c_1$ and $c_2$ are very small.
Actually, these quantities are not dimensionless.
According to the approach of Ref. \cite{steinhoff}, they can be identified with tidal deformation parameters which are usually made dimensionless through the radius of the body (see also Ref. \cite{damnag}).

\subsection{The case of spin-induced quadrupole tensor}

\begin{figure} 
\typeout{*** EPS figure 3}
\begin{center}
\includegraphics[scale=0.4]{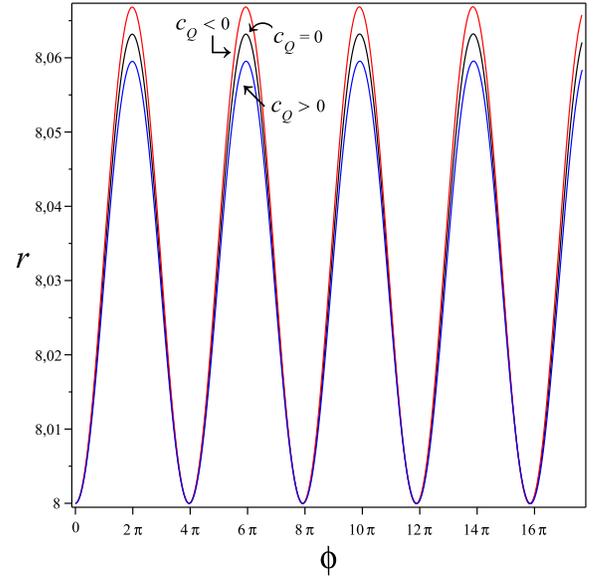}
\end{center}
\caption{The behavior of the radial coordinate as a function of the azimuthal coordinate is shown in the case of a spin-induced quadrupole tensor given by Eq. (\ref{Jspininduced}).
The set of MPD equations is numerically solved with the same initial conditions as in Fig. \ref{fig:1} and with the a value $s/(m_0M)=0.01$ of the spin parameter.
The curves correspond to different values $C_Q=[-100,0,100]$ of the quadrupole parameter (exaggerated to better show the effect).
}
\label{fig:3}
\end{figure}

Let us consider the case in which the quadrupole tensor is completely determined by the spin structure of the body as in Eq. (\ref{Jspininduced}).
The quadrupole force has the same form as in Eq. (\ref{Fquadframeu}) with 
\begin{eqnarray}
F_{\rm (quad)}^1&=&10\frac{N}{r}\zeta_K^2\frac{s^2}{m}C_Q\gamma_u^2\sin\alpha_u\nonumber\\
&&\times\left[\nu_u^2\cos^2\alpha_u-\frac15(1+2\nu_u^2)\right]\,,\nonumber\\
F_{\rm (quad)}^2&=&10\frac{N}{r}\zeta_K^2\frac{s^2}{m}C_Q\gamma_u^3\cos\alpha_u\nonumber\\
&&\times\left[\nu_u^2\cos^2\alpha_u-\frac15(1+4\nu_u^2)\right]\,.
\end{eqnarray}
and the torque is given by Eq. (\ref{torqueriem}) with 
\beq 
{\mathcal E}(u)_{\rm (quad)}=
-2\zeta_K^2\frac{s^2}{m}C_Q\gamma_u\nu_u\sin2\alpha_u(\omega^1-\gamma_u\tan\alpha_u\omega^2)\,.
\eeq
The evolution equations for $m$, $\nu_u$ and $\alpha_u$ have the same form as in Eq. (\ref{setfin}), whereas the spin invariant turns out to be constant because ${\mathcal B}(u)_{\rm (quad)}\equiv0$ in this case.
The numerical integration of the equations of motion with the same initial conditions as before, i.e., with the center of mass line initially aligned with a stable equatorial circular geodesic, gives again an oscillatory behavior of the radial coordinate.
The effect of the quadrupolar structure of the body is the increase/decrease of the oscillation amplitude for negative/positive values of the quadrupole parameter $C_Q$ (see Fig. \ref{fig:3}).


\subsection{The limiting case of vanishing quadrupole}

When all quadrupole functions vanish the MPD set of equations (\ref{setfin}) reduces to
\begin{eqnarray} 
\label{setfinsolospin}
\frac{\rmd m}{\rmd \tau} &=& 
F_{\rm (spin)}^0\nonumber\\
&=&3\frac{M}{r^3}\gamma\gamma_u^2 s \nu_u\sin\alpha_u(\nu\cos\alpha-\nu_u\cos\alpha_u)\,,\nonumber\\
\frac{\rmd \alpha_u}{\rmd \tau} &=& 
\frac{N}{r}\gamma\nu\sin\alpha 
-\frac{M}{r^2N}\frac{\gamma}{\nu_u}\sin\alpha_u\nonumber\\
&&-\frac{1}{m\gamma_u\nu_u} F_{\rm (spin)}^1  \,,\nonumber\\
\frac{\rmd \nu_u}{\rmd \tau} &=& 
-\frac{M}{r^2N}\frac{\gamma}{\gamma_u^2}\cos\alpha_u
+\frac{1}{m\gamma_u^2} F_{\rm (spin)}^2  \,,\nonumber\\
\frac{\rmd s}{\rmd \tau} &=&0\,.
\end{eqnarray}
Furthermore, since $A=0=C$ Eqs. (\ref{nuandalpha}) simplify to  
\beq
\label{nuandalphasolospin}
\tan\alpha=\frac{B}{D}\,, \qquad
\gamma=\frac{1}{\sqrt{1-B^2-D^2}}\,,
\eeq
with $B$ and $C$ still given by Eqs. (\ref{ABC_etc}), implying that 
\begin{eqnarray}
\nu&=&\sqrt{B^2+D^2}\nonumber\\
&=&\nu_u \left[\sin^2\alpha_u\left(\frac{m^2+2s^2\zeta_K^2}{m^2-s^2\zeta_K^2}\right)^2+\cos^2\alpha_u\right]^{1/2}\,,\nonumber\\
\end{eqnarray}
and
\begin{eqnarray}
\nu\sin\alpha&=&B=\nu_u\sin\alpha_u\frac{m^2+2s^2\zeta_K^2}{m^2-s^2\zeta_K^2}\,,\nonumber\\
\nu\cos\alpha&=&D=\nu_u\cos\alpha_u\,.
\end{eqnarray}
The latter equation implies that $F_{\rm (spin)}^0=0$, so that the mass $m$ turns out to be constant along the path as from the first equation of Eqs. (\ref{setfinsolospin}).

\section{Deviation from a circular geodesic}
\label{pertsol}

In order to avoid backreaction effects (which are expected, for example, in regions where the gravitational field is highly inhomogeneous, as noticed  in Ref. \cite{Semerak}), implicit in the MPD model is the requirement that the structure of the body should produce very small deviations from geodesic motion in the sense that the natural length scales associated with the body, i.e., the \lq\lq bare'' mass $m_0$, the spin length $|S^a|/m_0$ and the quadrupolar lengths $(|Q(u)_{ab}|/m_0)^{1/2}$, $(|W(u)_{ab}|/m_0)^{1/2}$ and $(|M(u)_{ab}|/m_0)^{1/2}$, must be small enough if compared with the length scale associated with the background curvature.
Therefore, it seems reasonable to introduce the conditions of ``small spin" and ``small quadrupole" from the very beginning, resulting in a simplified set of linearized differential equations which can be easily integrated as shown below.
This approach allows an analytic discussion of the problem in complete generality in this limit, which can be compared with the numerical solution of the previous section to the full set of nonlinear equations.

Let us solve the system of MPD equations perturbatively, by assuming that $U$ be tangent to a geodesic circular orbit in the equatorial plane at radius $r=r_0$ for vanishing spin and quadrupole. 
The associated 4-velocity is given by Eq. (\ref{Ugeocirc}), where the quantities $\gamma_K$, $\nu_K$, $\zeta_K$ defined in Eq. (\ref{kepler}) are understood to be evaluated at $r=r_0$ and a positive (negative) sign corresponds to co-rotating (counter-rotating) orbits with respect to increasing values of the azimuthal coordinate $\phi$. 
We further assume just as in the previous section that the spin vector be orthogonal to the equatorial plane where the motion is confined and that the quadrupole tensor be represented by two independent components only in the $u$-frame, i.e., $X(u)_{11}$ and $X(u)_{22}$, in order to make the comparison. 
It is useful to introduce the following parameters 
\beq
{\hat s}=\frac{s}{m_0}\zeta_K\,,\qquad
{\hat X}_{ab}=\frac{X(u)_{ab}}{m_0}\zeta_K^2\,,
\eeq
associated with spin and quadrupole respectively, which will be taken to be much smaller than unity as smallness indicators (i.e., $|{\hat s}|\ll1$ and $|{\hat X}_{ab}|\ll1$).
Note that in this approximation scheme quantities which are linear in ${\hat s}$ will be considered as ``first order," whereas quantities which are linear in ${\hat X}_{ab}$ will be considered as ``second order." 
Therefore, the spin will contribute both to the first order and to the second, whereas the quadrupole to the second order only. 

Let us look for solutions of the form 
\beq
x^\alpha(\tau)=  x^\alpha_{\rm (geo)}(\tau) +x^\alpha_{(1)}(\tau)+x^\alpha_{(2)}(\tau)
\eeq
for the resulting path of the extended body, where the subscripts indicate the order of approximation.
Similarly for the other quantities involved, i.e., $m$, $\nu$, $\alpha$, $\nu_u$, $\alpha_u$, we have
\begin{eqnarray} 
m&=&m_0+m_{(2)}(\tau)\,,\nonumber\\
\nu&=&\pm\nu_K+\nu_{(1)}(\tau)+\nu_{(2)}(\tau)\,,\nonumber\\
\alpha&=&\frac{\pi}{2}+\alpha_{(1)}(\tau)+\alpha_{(2)}(\tau)\,,\nonumber\\
\nu_u&=&\pm\nu_K+\nu_{u(1)}(\tau)+\nu_{u(2)}(\tau)\,,\nonumber\\
\alpha_u&=&\frac{\pi}{2}+\alpha_{u(1)}(\tau)+\alpha_{u(2)}(\tau)\,.
\end{eqnarray}
Note that the spin invariant $s$ remains constant along the path and the mass varies only to second order due to the quadrupole, as expected.

We are interested in solutions which describe deviations from geodesic motion due to both the spin-curvature force and the quadrupolar force. 
Hence, we choose initial conditions so that the 4-velocity is tangent to the circular geodesic, i.e., 
\beq
x^{\alpha}_{(1)}(0)=0=\frac{\rmd x^{\alpha}_{(1)}(0)}{\rmd\tau}\,, \qquad
x^{\alpha}_{(2)}(0)=0=\frac{\rmd x^{\alpha}_{(2)}(0)}{\rmd\tau}\,,
\eeq
and similarly for the remaining first order and second order quantities.

To first order we have:
\begin{eqnarray} 
\label{eqsord1}
 \frac{\rmd t_{(1)}}{\rmd \tau} &=& 
 \frac{\gamma_K \nu_K^2}{r_0 \zeta_K} \left(  -\frac{\nu_K}{r_0} r_{(1)} \pm \gamma_K^2 \nu_{(1)} \right) \,,
\nonumber\\
 \frac{\rmd r_{(1)}}{\rmd \tau} &=&
  \mp r_0\zeta_K\gamma_K \alpha_{(1)}\,,
\nonumber\\
 \frac{\rmd \phi_{(1)}}{\rmd \tau} &=&
 \pm \frac{\gamma_K }{r_0} \left(  -\frac{\nu_K}{r_0} r_{(1)} \pm\gamma_K^2 \nu_{(1)} \right)
=\pm\frac{\zeta_K}{\nu_K^2} \frac{\rmd t_{(1)}}{\rmd \tau}\,, 
\nonumber\\
 \frac{\rmd \nu_{(1)}}{\rmd \tau} &=&
 \frac{\nu_K\zeta_K}{\gamma_K}\alpha_{(1)}\,, 
\nonumber\\
 \frac{\rmd \alpha_{(1)}}{\rmd \tau} &=&
-2\frac{\gamma_K\zeta_K}{\nu_K}\nu_{(1)}
\mp\frac{\gamma_K\nu_K^2}{r_0^3\zeta_K}r_{(1)}
-3\gamma_K\zeta_K{\hat s}\,,\nonumber\\
\end{eqnarray}
with
\beq
\nu_{u(1)}=\nu_{(1)}\,,\qquad
\alpha_{u(1)}=\alpha_{(1)}\,,
\eeq
and the circular geodesic is described by the equations
\begin{eqnarray} 
t_{\rm (geo)} &=&  \Gamma_K \tau +t_0\,,\qquad
r_{\rm (geo)} = r_0\,,\nonumber\\
\theta_{\rm (geo)} &=&\frac{\pi}{2}\,,\qquad
\phi_{\rm (geo)} =  \pm\Omega_{\rm(orb)} \tau + \phi_0\,, 
\end{eqnarray}
with
\beq
\Omega_{\rm(orb)}=\frac{\gamma_K \nu_K}{r_0}= \frac{1}{r_0}\sqrt{\frac{M}{r_0-3M}}
\,.
\eeq

The solutions to the equations (\ref{eqsord1}) are given by
\begin{eqnarray} 
\label{solord1}
t_{(1)}&=& 
 \frac{6\gamma_K^3 \nu_K^3\zeta_K}{r_0\Omega_{\rm(ep)}^3}{\hat s} (\sin\Omega_{\rm(ep)}\tau-\Omega_{\rm(ep)}\tau) \,,
\nonumber\\
r_{(1)}&=&
  \mp \frac{3r_0\zeta_K^2\gamma_K^2}{\Omega_{\rm(ep)}^2}{\hat s} (\cos\Omega_{\rm(ep)}\tau-1)\,,
\nonumber\\
\phi_{(1)}&=&\pm\frac{\zeta_K}{\nu_K^2} t_{(1)}\,, 
\nonumber\\
\nu_{(1)}&=&
\frac{3\nu_K\zeta_K^2}{\Omega_{\rm(ep)}^2}{\hat s} (\cos\Omega_{\rm(ep)}\tau-1)\,,
\nonumber\\
\alpha_{(1)}&=&
-\frac{3\zeta_K\gamma_K}{\Omega_{\rm(ep)}}{\hat s} \sin\Omega_{\rm(ep)}\tau\,,
\end{eqnarray}
where
\beq
\Omega_{\rm(ep)} \equiv \sqrt{\frac{M (r_0-6M)}{r_0^3 (r_0 -3M)}}
\eeq
is the well known epicyclic frequency.

Therefore, the first order solution is characterized by an oscillatory behavior of the radial component about the geodesic orbit in a circular ring either inside or outside the geodesic radius depending on the relative sign of the vertical component of the spin and the orbital velocity.  
The azimuthal motion also oscillates around the geodesic value with the same frequency characterizing the radial motion, apart from a secular drift which occurs at slightly different speeds for the inner and outer radial oscillations (see also Ref. \cite{spin_dev_schw}).

To second order we have:
\begin{widetext} 
\begin{eqnarray} 
\label{eqsord2}
 \frac{\rmd t_{(2)}}{\rmd \tau} &=&
 -\frac{\gamma_K \nu_K^3}{r_0^2 \zeta_K} r_{(2)} 
 \pm\frac{\gamma_K^3 \nu_K^2}{r_0 \zeta_K} \nu_{(2)} 
 +\frac{9}{2}{\hat s}^2 \frac{\gamma_K^5\nu_K^5\zeta_K}{r_0^3\Omega_{\rm(ep)}^4}(3+r_0^2\zeta_K^2)(\cos\Omega_{\rm(ep)}\tau-1)^2 \,,
\nonumber\\
 \frac{\rmd r_{(2)}}{\rmd \tau} &=&
  \mp r_0\zeta_K\gamma_K \alpha_{(2)}
+9{\hat s}^2 \frac{\gamma_K^2r_0\zeta_K^4\nu_K^5}{\Omega_{\rm(ep)}^3}(\cos\Omega_{\rm(ep)}\tau-1)\sin\Omega_{\rm(ep)}\tau  \,,
\nonumber\\
 \frac{\rmd \phi_{(2)}}{\rmd \tau} &=&\pm\frac{\zeta_K}{\nu_K^2} \frac{\rmd t_{(2)}}{\rmd \tau} 
 \pm9{\hat s}^2
 \frac{\gamma_K^3\zeta_K^2\nu_K}{r_0\Omega_{\rm(ep)}^2}(\cos\Omega_{\rm(ep)}\tau-1)\cos\Omega_{\rm(ep)}\tau \,, 
\nonumber\\
 \frac{\rmd \nu_{(2)}}{\rmd \tau} &=&
 \frac{\nu_K\zeta_K}{\gamma_K}\alpha_{(2)} \mp9{\hat s}^2
\frac{\zeta_K^2\nu_K}{\Omega_{\rm(ep)}^3}[2\gamma_K^2\zeta_K^2(\cos\Omega_{\rm(ep)}\tau-1)-\Omega_{\rm(ep)}^2]\sin\Omega_{\rm(ep)}\tau \,, 
\nonumber\\
 \frac{\rmd \alpha_{(2)}}{\rmd \tau} &=&
-2\frac{\gamma_K\zeta_K}{\nu_K}\nu_{(2)}
\mp\frac{\gamma_K\nu_K^2}{r_0^3\zeta_K}r_{(2)}
\pm2\frac{\gamma_K}{r_0^2\zeta_K}(-3+7r_0^2\zeta_K^2){\hat X}_{11}
\mp2\zeta_K\gamma_K{\hat X}_{22}\nonumber\\
&&\pm3\gamma_K\zeta_K{\hat s}^2\left[
1+3\frac{\gamma_K^2\nu_K^2}{r_0^2\Omega_{\rm(ep)}^2}(\cos\Omega_{\rm(ep)}\tau-1)\left(
3-5r_0^2\zeta_K^2-\frac{\gamma_K^2\zeta_K^2\nu_K^2}{\Omega_{\rm(ep)}^2}(5-3r_0^2\zeta_K^2)(\cos\Omega_{\rm(ep)}\tau-1)
\right)
\right]\,,\nonumber\\
\end{eqnarray}
\end{widetext} 
with
\begin{eqnarray} 
\nu_{u(2)}&=&\nu_{(2)}\mp\nu_K(3{\hat s}^2+8{\hat X}_{11}+4{\hat X}_{11})\,,\nonumber\\
\alpha_{u(2)}&=&\alpha_{(2)}\,.
\end{eqnarray}

The solutions to the equations (\ref{eqsord2}) are given by
\begin{eqnarray} 
\label{solord2}
t_{(2)}&=& D_1\sin\Omega_{\rm(ep)}\tau+D_2\sin2\Omega_{\rm(ep)}\tau+D_3\tau\cos\Omega_{\rm(ep)}\tau\nonumber\\
&&+D_4\tau\,, \nonumber\\
r_{(2)}&=& C_1(\cos\Omega_{\rm(ep)}\tau-1)+C_2(\cos2\Omega_{\rm(ep)}\tau-1)\nonumber\\
&&+C_3\tau\sin\Omega_{\rm(ep)}\tau\,, \nonumber\\
\phi_{(2)}&=& E_1\sin\Omega_{\rm(ep)}\tau+E_2\sin2\Omega_{\rm(ep)}\tau+E_3\tau\cos\Omega_{\rm(ep)}\tau\nonumber\\
&&+E_4\tau\,, \nonumber\\
\nu_{(2)}&=& A_1(\cos\Omega_{\rm(ep)}\tau-1)+A_2(\cos2\Omega_{\rm(ep)}\tau-1)\nonumber\\
&&+A_3\tau\sin\Omega_{\rm(ep)}\tau\,, \nonumber\\
\alpha_{(2)}&=& B_1\sin\Omega_{\rm(ep)}\tau+B_2\sin2\Omega_{\rm(ep)}\tau+B_3\tau\cos\Omega_{\rm(ep)}\tau\,,\nonumber\\
\end{eqnarray}
where the integration constants are listed in Appendix \ref{const} (see Eq. (\ref{solord2coeffs})).

Therefore, the second order solutions are still oscillatory as those of first order, but with two different frequencies, the epicyclic one and twice it.
Furthermore, the second order quantities all contain secular terms which increase with proper time.
Those terms are responsible for the band wherein the motion is confined to widen after each revolution, so confirming the results of the previous section where this same feature was found to occur for very small values of both the spin and quadrupole parameters by solving numerically the full set of nonlinear equations. 

Finally, the perturbed trajectory $r(\phi)$ is given by
\begin{eqnarray} 
r(\phi)&=&r_0\mp3\frac{r_0\zeta_K^2\gamma_K^2}{\Omega_{\rm(ep)}^2}{\hat s}(\cos\psi-1)
+[C_1(\cos\psi-1)\nonumber\\
&&+{\tilde C_2}(\cos2\psi-1)
+{\tilde C_3}\psi\sin\psi]\,,
\end{eqnarray}
where $\psi=[{\Omega_{\rm(ep)}}/{\Omega_{\rm(orb)}}](\phi-\phi_0)$ and ${\tilde C_2}$ and ${\tilde C_3}$ are given in Eq. (\ref{solord2coeffs2}).

\section{Concluding remarks}

We have investigated the dynamics of extended bodies endowed with intrinsic spin and quadrupole moment in the Schwarzschild spacetime according to the Mathisson-Papapetrou-Dixon model, extending previous works.
The motion of the center of mass line used for the multipole reduction has been assumed to be confined on the equatorial plane, the spin vector being orthogonal to it.
This is the simplest choice and is also useful for applications to astrophysical systems, like neutron stars or binary pulsar systems orbiting the Galactic Center.
In order to study the effect of the mass quadrupole moment of an extended body on its motion we have considered the case in which the quadrupole tensor is completely specified by two independent components only.
Imposing such a condition is not so restrictive. In fact, it does not affect the main features of the underlying physics and can be easily relaxed.
Furthermore, we have fixed the freedom in determining the components of the quadrupole tensor due to the lack of evolution equations in the MPD model by assuming them to be constant with respect to the frame associated with the 4-momentum of the body itself; in this sense, they are known as intrinsic properties of the matter under consideration. 
We have also discussed the possibility to construct the quadrupole tensor in a different way, e.g., by assuming it to be directly related to the Riemann tensor, having the same symmetry properties, or even by deriving it from a suitable Lagrangian within an action principle formulation of the model equations available from recent literature.
However, these latter approaches do not seem to have a particular physical meaning in the context of the MPD model: here in fact one expects the quadrupole tensor to represent the moving matter only, with no a priori relations with the background in which the motion takes place.

We have found that the presence of the quadrupole significantly changes the features of the motion with respect to the case of a purely spinning body.
In fact, both the mass and the spin invariant are no longer constant along the path, as a general result.
Furthermore, the quadrupolar structure of the body is responsible for the onset of spin angular momentum, if the body is initially not spinning, a fact that does not seem to have received enough attention in the literature.
To illustrate these general properties we have numerically integrated the full set of MPD equations for different choices of parameters and initial conditions. 
In particular, we have considered the case in which the orbit is initially tangent to a stable equatorial circular geodesic for a particle without structure.
We have found that in general the trajectory of the extended body oscillates filling a nearly circular corona of fixed width (depending on the chosen values of the spin and quadrupole parameters) around the geodesic path, similarly to the case of a purely spinning particle.
The situation turns out to be better elucidated especially when the characteristic length scales associated with the spin and quadrupole are taken to be very small with respect to the background curvature characteristic length, which is the limit of validity of the MPD model.
In fact, in this case the thickness of the region wherein the body moves slightly increases after each revolution.
A confirmation to this result comes from the analytic solution of the MPD equations for small values of both spin and quadrupole parameters, showing an oscillatory behavior of the orbit characterized by the occurrence of a secular increase of the bandwidth.
This effect is obviously strongly suppressed if initially the body is also endowed with spin, whose value almost determines the oscillation amplitude.

\appendix

\section{Perturbative solution}
\label{const}

We list below the integration constants of the perturbative solution of Section \ref{pertsol}:
\begin{eqnarray} 
\label{solord2coeffs}
A_1&=&
\pm\frac32\frac{\gamma_K^4\nu_K^5\zeta_K^2}{r_0^4\Omega_{\rm(ep)}^6}\left(55-162r_0^2\zeta_K^2-69\frac{r_0^4\zeta_K^4}{\nu_K^4}\right){\hat s}^2\nonumber\\
&&\pm2\frac{\nu_K}{r_0^2\Omega_{\rm(ep)}^2}(3-7r_0^2\zeta_K^2){\hat X}_{11}
\pm2\frac{\nu_K\zeta_K^2}{\Omega_{\rm(ep)}^2}{\hat X}_{22}\,,\nonumber\\
A_2&=&
\mp\frac98\frac{\gamma_K^4\nu_K^5\zeta_K^2}{r_0^4\Omega_{\rm(ep)}^6}\left(11-18r_0^2\zeta_K^2-17\frac{r_0^4\zeta_K^4}{\nu_K^4}\right){\hat s}^2\,,\nonumber\\
A_3&=&
\pm9\frac{\gamma_K^4\nu_K^3\zeta_K^4}{r_0^2\Omega_{\rm(ep)}^5}(1-9r_0^2\zeta_K^2){\hat s}^2\,,\nonumber\\
B_1&=&
-\frac{\gamma_K\Omega_{\rm(ep)}}{\nu_K\zeta_K}A_1\nonumber\\
&&\pm\frac92\frac{\gamma_K^5\zeta_K}{\nu_K^4r_0^4\Omega_{\rm(ep)}^5}\left(17-50r_0^2\zeta_K^2-21\frac{r_0^4\zeta_K^4}{\nu_K^4}\right){\hat s}^2
\,,\nonumber\\
B_2&=&
\pm\frac94\frac{\gamma_K^5\nu_K^4\zeta_K}{r_0^4\Omega_{\rm(ep)}^5}\left(3-2r_0^2\zeta_K^2-5\frac{r_0^4\zeta_K^4}{\nu_K^4}\right){\hat s}^2\,,\nonumber\\
B_3&=&
A_3\frac{\gamma_K\Omega_{\rm(ep)}}{\nu_K\zeta_K}\,,\nonumber\\
C_1&=&
\pm\frac{\gamma_Kr_0\zeta_K}{\Omega_{\rm(ep)}}B_1
+18\frac{\gamma_K^6\zeta_K^4\nu_K^4}{r_0\Omega_{\rm(ep)}^6}{\hat s}^2\,,\nonumber\\
C_2&=&
-\frac92\frac{\gamma_K^6\nu_K^2\zeta_K^4}{r_0\Omega_{\rm(ep)}^6}(1-7r_0^2\zeta_K^2){\hat s}^2\,,\nonumber\\
C_3&=&
\mp A_3\frac{r_0\gamma_K^2}{\nu_K}\,,\nonumber\\
D_1&=&
\mp2\frac{\gamma_K^2\nu_K^3}{r_0\Omega_{\rm(ep)}^2}B_1
-9\frac{\gamma_K^7\zeta_K\nu_K^7}{r_0^5\Omega_{\rm(ep)}^7}(3-10r_0^2\zeta_K^2){\hat s}^2\,,\nonumber\\
D_2&=&
-\frac94\frac{\gamma_K^7\nu_K^7\zeta_K}{r_0^5\Omega_{\rm(ep)}^7}\left(3+2r_0^2\zeta_K^2-7\frac{r_0^4\zeta_K^4}{\nu_K^4}\right){\hat s}^2\,,\nonumber\\
D_3&=&
\mp2A_3\frac{\gamma_K^3\nu_K^2}{r_0\zeta_K\Omega_{\rm(ep)}}\,,\nonumber\\
D_4&=&
-\Omega_{\rm(ep)}D_1
-9\frac{\gamma_K^7\zeta_K^3\nu_K^7}{r_0^3\Omega_{\rm(ep)}^6}(7-22r_0^2\zeta_K^2){\hat s}^2\,,\nonumber\\
E_1&=&
2\frac{\gamma_K^3}{r_0\Omega_{\rm(ep)}}A_1\nonumber\\
&&\mp108\frac{\gamma_K^7\zeta_K^2}{\nu_K^3r_0^5\Omega_{\rm(ep)}^7}\left(1-3r_0^2\zeta_K^2-\frac{r_0^4\zeta_K^4}{\nu_K^4}\right){\hat s}^2
\,,\nonumber\\
E_2&=&
-\frac{E_3}{\Omega_{\rm(ep)}}
\pm\frac{13}{4}\frac{\gamma_K^7\zeta_K^4\nu_K^3}{r_0^3\Omega_{\rm(ep)}^7}{\hat s}^2\,,\nonumber\\
E_3&=&
-2A_3\frac{\gamma_K^3}{r_0\Omega_{\rm(ep)}}\,,\nonumber\\
E_4&=&
-\Omega_{\rm(ep)}E_1
\mp\frac92\frac{\gamma_K^7\zeta_K^4\nu_K^3}{r_0^3\Omega_{\rm(ep)}^6}(1+4r_0^2\zeta_K^2){\hat s}^2
\,,
\end{eqnarray}
and
\begin{eqnarray}
\label{solord2coeffs2}
{\tilde C_2}&=&
C_2+9\frac{\gamma_K^5\nu_K\zeta_K^4}{\Omega_{\rm(orb)}\Omega_{\rm(ep)}^4}{\hat s}^2
\,,\nonumber\\
{\tilde C_3}&=&
\frac{C_3}{\Omega_{\rm(ep)}}+18\frac{\gamma_K^5\nu_K\zeta_K^4}{\Omega_{\rm(orb)}\Omega_{\rm(ep)}^4}{\hat s}^2
\,.
\end{eqnarray}

\section*{Acknowledgements}
The authors are indebted to Prof. T. Damour for useful discussions.

\end{document}